%% file: Author_tex.tex
\documentclass[]{rsos}

\usepackage[titletoc,title]{appendix}
\usepackage{amsmath,amssymb}

\usepackage{rotating}
\usepackage{pdflscape}
\usepackage{bm}

\newcommand\citep{\cite}
\newcommand\citet{\cite}

\usepackage{comment}

\begin{document}

\vspace{0.5cm}

\noindent{\Large{\bfseries Estimating the drivers of urban economic complexity and their connection to economic performance}}
\vspace{0.5cm}

\noindent{\Large{ Andres Gomez-Lievano${}^{1,2}$ and Oscar Patterson-Lomba${}^{2}$}}\\
\small{${}^{1}$ Growth Lab, Harvard University, Cambridge MA, USA}\\
\small{${}^{2}$ Analysis Group, Inc., Boston MA, USA}

\section*{Abstract}
Estimating the capabilities{, or inputs of production,} that drive and constrain the economic development of urban areas has remained a challenging goal. We posit that capabilities are instantiated in the complexity and sophistication of urban activities, the knowhow of individual workers, and the city-wide collective knowhow. We derive a model that indicates how the value of these three quantities can be inferred from the probability that an individual in a city is employed in a given urban activity. We illustrate how to estimate empirically these variables using data on employment across industries and metropolitan statistical areas in the US. { We then show how the functional form of the probability function derived from our theory is statistically superior when compared to competing alternative models, and that it explains well-known results in the urban scaling and economic complexity literature. Finally, we} show how the quantities are associated with metrics of economic performance, suggesting our theory can provide testable implications for why some cities are more prosperous than others.

\vspace{0.5cm}

\noindent{\emph{Subject Category: Science, society and policy}\\
\emph{Subject Areas: Mathematical modelling, complexity, biogeography}\\
\emph{Keywords: Economic complexity, collective know-how, industry complexity, urban employment}}

\vspace{0.5cm}

\section{Introduction} 

{Fine-grained representations of the economy of countries, regions and cities in terms of what they produce, and the industries they have, have revealed that the location of economic activities across places is not random \citep{ellison1997geographic,HidalgoHausmann2009,caldarelli2012network,saracco2015randomizing,straka2017grand,hidalgo2021economic}. 
Instead, activities tend to locate in cities of different sizes depending on the number of inputs they require \citep{gomez2016explaining,balland2018complex} and co-locate with other activities that require a similar set of inputs \citep{porter2003economic,HidalgoEtAl2007,boschma2011emerging,boschma2017relatedness,HidalgoEtAl2018PoR,diodato2018industries}. 
While there is an extensive research program in the traditional economic growth and development literature for understanding how specific inputs, such as natural resources, public goods, labor, physical and human capital, institutions, or amenities, determine the choices of workers and firms for where to locate and what to produce \citep{fujita1999spatial}, there is less research on more agnostic models that do not specify \emph{a priori} what those inputs are. The latter has been a big part of the research program in the interdisciplinary field of economic complexity \citep{hidalgo2021economic}, and several dimensionality reduction algorithms have been applied to data to extract `metrics of complexity' which quantify the availability and sophistication of inputs present in an economy \citep{HidalgoHausmann2009,TacchellaCristelliCaldarelliEtAl2012SciRep,tacchella2013economic,cristelli2013measuring,balland2017geography,brummitt2020machine,mealy2019interpreting,sciarra2020reconciling,hidalgo2021economic}. 

The complexity metrics computed from these algorithms were originally developed to explain differences in wealth and trade patterns across countries \citep{HidalgoHausmann2009,tacchella2013economic,tacchella2018dynamical}, and have since been shown to be highly correlated with both the levels and the change of aggregate economic output at higher geographical resolutions, such as regions and cities. For example, economic complexity metrics were recently applied to US regions and cities \citep{fritz2021economic}, prefectures in Japan \citep{chakraborty2020economic}, provinces in China \citep{gao2018quantifying}, Indian states \citep{sahasranaman2020economic}, Mexican states \citep{hausmann2021place}, and Colombian cities \citep{OClery2018}, and shown to explain different aspects of their economic growth (see \citep{gao2019computational,hidalgo2021economic} for a review of the literature). 

So far, however, these dimensionality reduction algorithms have been difficult to connect with theoretical models of how economies work (see  \citep{schetter2019structural,bustos2020production}). That is, there is no assurance that the manipulations of the data that these algorithms conduct actually quantify what they claim to quantify: the number of capabilities available in an economy, or required by an economic activity. For this reason, their use and interpretation has remained highly contested \citep{mariani2015measuring,mealy2019interpreting,morrison2017economic}. Ensuring that the mathematical foundations of economic complexity metrics are robust, that their method of estimation is reliable, and that they stand as meaningful economic indicators is key to understanding cities as complex systems, and acting upon them as such. In this work, we present a novel first-principles model that aims to address these shortcomings. That is, we present how a simple mathematical model can characterize the way inputs combine in cities to generate output (without making strong assumptions about what these inputs may be), how complexity variables then emerge and a method to estimate them, and how they correlate with measures of urban economic performance.}

\section{Model of urban economic complexity}
{The model we propose seeks to understand the production of units of output in cities. For illustration purposes, we will consider employment as the output of interest, but other forms of production such as the creation of firms, the production of patents, or occurrence of crime, could have been considered as well.} 

In this context, we propose a ``production recipes'' approach to understand the patterns of employment probabilistically \cite{AuerswaldEtAl2000}. This approach suggests that questions like \emph{why does a given industry employ more workers per capita in some cities than in others?} or \emph{why in a given city do some industries employ more workers than others?} should be framed as questions about the mathematical shape of the urban production function across different types of phenomena. The production recipes approach is different from traditional economic models because it assumes that the statistics observed in economic data are more a function of how inputs combine rather than what those inputs are. The simplicity of this probabilistic approach emphasizes that urban production processes go beyond economic processes, which is why it can include phenomena like crime and disease.

Several models and explanations have been proposed to make sense of the differential shares of employment across industries. We argue that our proposed model offers a simple alternative that, for example, does not need assumptions about prices and markets to account for the empirical patterns observed (see \citep{ottaviano2004agglomeration} for a review of the literature). Two empirical patterns suggest that the mechanisms go beyond economics, and thus support our choice for a production recipes approach: the matrix of cities and industry presence displays ``nestedness'' \cite{BustosEtAl2012}, and the employment is associated with a superlinear function of population size \cite{YounEtAl2015Universality}. That is, ``nestedness'' whereby larger cities have most industries (including those present in the least diversified cities), and urban scaling whereby larger cities have, on average, more absolute employment in any given industry, relative to small cities. The increase in employment $Y_{c,f}$ in city $c$ in an industry $f$ with city size follows a power-law function, $Y_{c,f}\propto N_c^\beta$ \cite{YounEtAl2015Universality}. The nestedness and the power-law are also observed in other urban phenomena such as cases of crime, infectious disease prevalence, educational attainment, and technological innovation \cite{BettencourtLoboStrumsky2007,BettencourtPNAS2007,BettencourtPLOS2010,GomezLievanoYounBettencourt2012,Patterson2015STDs,rocha2015non,alves2015scale,gomez2016explaining}. The nestedness and the power-law relationships across such diverse set of phenomena beyond economics are both key empirical observations that suggest that a general mechanism must be at work (see \citep{DavisDingel2014NBER,schetter2019quality} for Ricardian models attempting to explain the nestedness). {In Sections~\ref{sec_implications_us} and \ref{sec_implications_nest} we show how scaling laws and nestedness emerge from the model we propose.}

The theory we apply here is based on three assumptions. First, industries are defined as different conjunctions of complementary factors, as it has been shown in \cite{neffke2019value}. Complementarity implies that, for a unit of employment to be created in an industry, all factors must be simultaneously present. More complex industries are those that require more underlying causal factors. Second, cities possess a multiplicity of these factors\footnote{In \citep{gomez2016explaining} we developed a model in which factors are acquired through a stochastic process of accumulation, but here we simply take the factors present in a city as a given}. And third, each person in the city have an (individual) ability to get their own factors. In general, we refer to these factors as ``capabilities'', and we use these terms interchangeably. We claim that the model reveals a simple methodology to estimate these underlying complexity metrics that allows us to understand the differences in average wages and average size of establishments across cities and industries. 

{Since the model is built upon the concept of a `capability', it is worth briefly discussing what capabilities stand for. The idea is that capabilities are inputs of production, but getting the right capabilities may require different instances of knowledge. Hence, a capability can be acquired either by know-what (e.g., knowing which specific component is needed for a machine to operate), know-why (knowing the causal mechanisms that are critical for the machine to work), know-who (knowing who has a specific expertise to fill in a specific technique), and so on. In general, however, there are reasons to believe that the most determinant component of knowledge is tacit \citep{johnson2002all,kogut1992knowledge,coscia2020knowledge}. This type of knowledge is generally referred to as `knowhow'. Knowhow can be possessed by the individual in question, or distributed in the city and accessed by living in it and interacting with its infrastructure and its citizens \citep{coscia2020knowledge}. This is the reason we use `knowhow' as part of the description of the variables derived from the model. In our model, the person-specific  probability  of  possessing  any  one  capability ($s_i$) may be associated with occupation-specific and industry-specific knowhow carried by  individuals, while the city-specific probability of possessing any one capability ($r_c$) may be associated with distributed knowhow, specific to a location, collectively possessed by the city (see, however, \citep{jara2018role} for a study in which each of these types of knowledge is claimed to be internalized by workers and shown to affect patterns of economic diversification; for a review of the factors affecting diversification see \citep{content2016related}).}

\begin{figure}[t!]
	\centering
		\includegraphics[width=0.8\textwidth, trim = 0in 0in 0in 0in]{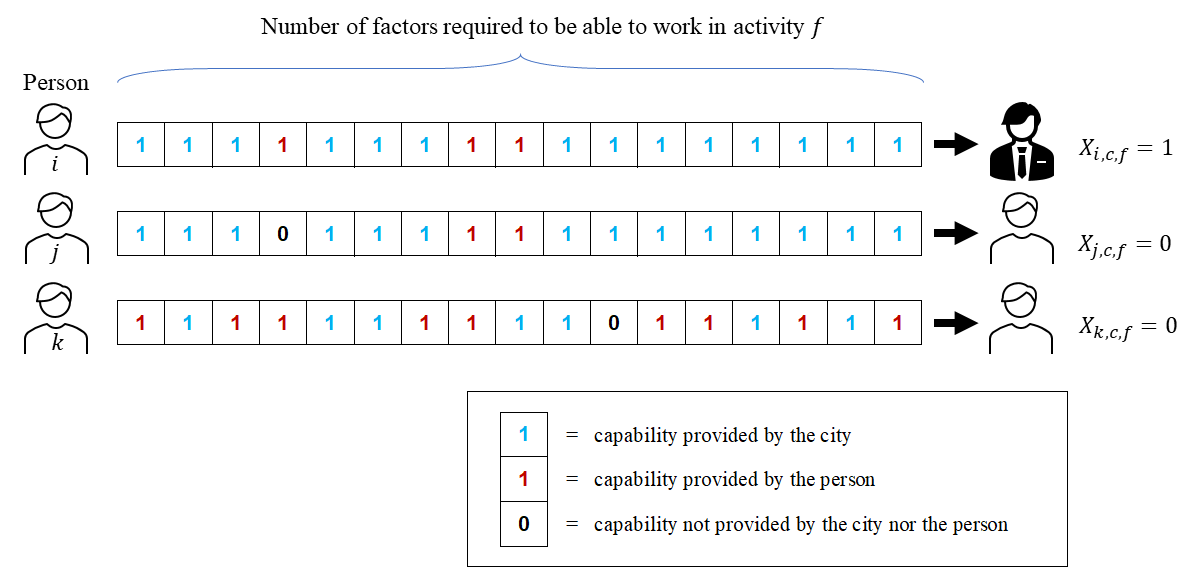}
		\caption{{\bfseries Sketch of the model.} Three different individuals $i$, $j$ and $k$ in city $c$ and activity $f$. Individuals get their capabilities to work in $f$ either from their own effort (red 1's) or from the city (blue 1's). Note that individuals may differ in the way they get the capabilities. Only individual $i$ is able to get a job at activity $f$ in the city. Individuals $j$ and $k$ fail to get employed because they failed to get one of the capabilities. We quantify the number of factors required by activity $f$ by $M_f$, the likelihood of drawing a 1 from city $c$ by $r_c$, and the likelihood of drawing a 1 by the $i$th individual by $s_i$.} 
	\label{Fig_sketch}
\end{figure}
{Figure~\ref{Fig_sketch} presents a conceptual sketch of how our model works, and shows that production is the process of combining a multiplicity of complementary inputs (for similar models, see \citep{kremer1993ring,Weitzman1998,HidalgoHausmann2009,HausmannHidalgo2011}).   
In the model, individuals will be indexed with the letter $i$, cities with $c$ and industries with $f$. With this setup and the conceptual sketch in Figure~\ref{Fig_sketch}, three fundamental variables emerge from the model:}
\begin{enumerate}
	\item $M_f$: This is the number of capabilities required to produce the industry-specific product $f$. We can refer to it as the intrinsic ``complexity'' of the industry.
	\item $s_i$: This is the person-specific probability of possessing any one capability. Accordingly, $s_i$ can be referred to as a measure of ``individual knowhow''.
	\item $r_c$: This is the city-specific probability of possessing any one capability. Accordingly, $r_c$ can be referred to as a measure of ``collective knowhow'' and represents a measure of input availability in the urban environment of city $c$.	
\end{enumerate}
{Importantly, these  parameters  will be estimated from data, rather than estimated based on \emph{ad-hoc} definitions. Specifically, using cross-sectional data on employment across cities and industries, we will make inferences about the current value of these hidden variables.}

To derive a connection between data and these variables we need to understand how, in this model, the recombination of complementary capabilities in cities gives rise to probabilities of employment. To that end, let the random variable $X$ be equal to one if the person is employed and zero if not. We will derive the probability, $p$, that someone is employed in a city in an industry as a function of the complexity of the industry complexity $M_f$, the person's individual knowhow $s_i$, and the city's collective knowhow $r_c$. We will not, however, model the dynamics of $M_f$, $s_i$, or $r_c$. In particular, introducing laws of agglomeration or congestion are left out for future research (see electronic supplementary material~B for a brief discussion).

The parameter $M_f$ represents the ``inherent complexity'' of the economic activity associated with the production of industry's product $f$. The more capabilities are needed, the larger the value of $M_f$, and the more complex the activity. {We emphasize that the model is fully agnostic as to what capabilities are involved in these activities. The complexity $M_f$ of an industry may not necessarily be associated with the number of employees required by a typical firm, nor with the level of schooling of the workers employed in $f$ (e.g., industries that tend to employ people with PhDs versus industries that employ low skilled workers). Thus, $M_f$ may ultimately quantify something intangible and difficult to measure directly in the real world.} For example, industries that use low human capital but that are intensive in physical capital may have high $M_f$ if they require the coordination of a large set of inputs and materials to be transformed with the use of sophisticated machinery in a complex production process. 

The parameter $s_i$ represents how many capabilities the individual $i$ can acquire on their own. Specifically, it is the probability that she can get \emph{any} of the $M_f$ capabilities required by the typical business in industry $f$. This probability can be interpreted as a measure of their individual knowhow. Since capabilities are supposed to vary qualitatively from one another, $s_i$ is meant to capture the \emph{breadth} rather than the \emph{depth} of knowledge. It is about how many different things they could know how to do by themselves. The explicit inclusion of this parameter in the model represents an important contribution to the framework of economic complexity (cf. \cite{bustos2020production}). { Since capabilities represent units of knowledge that are interconnected, depth and breadth of knowledge (as measured by their level of schooling, or by the different things they knows how to do, respectively) may be both positively correlated with $s_i$.}

The parameter $r_c$ models the idea that the city provides workers with a variety of capabilities through exposure to other people, services and institutions within the urban milieu. Presumably, the bigger the city, the more diverse, and the more capabilities it can offer. Thus, $r_c$ is the \emph{probability} that the city $c$ provides \emph{any} capability.\footnote{We can imagine that the city has a ``field'' spread in the $xy$-coordinates, $r_c(x,y)$. This field is an abstraction of the urban milieu that represents the location-specific probability that the city provides one of the ingredients for phenomena to occur, and we can assume that people interact with it as they live and work in the city. It should capture the elements from all the types of urban interactions to which people are exposed through the social, economic and built environment. In this view, the city functions as though it is permeated across space by a ``cultural field'', and $r_c(x,y)$ quantifies the magnitude of the social, economic, and cultural repertoire available at a particular location. Locations where the value of the field is high have a high concentration of diverse urban factors.} 

By solving the model, we get that the probability that individual $i$ will be employed in industry $f$ given that they live in city $c$ can be written as:
\begin{equation}
	\Pr\{X_{i,c,f}=1\}=e^{-M_f (1-s_i)(1-r_c)}. \label{epercapita}
\end{equation}
See electronic supplementary material~A for details on the derivation of Equation~(\ref{epercapita}).

Our model shares many similarities with the O-ring model of production \citep{kremer1993ring}. As in the O-ring model, for example, employment is guaranteed only if all required capabilities (or ``tasks'') are combined. Our model differs from it, however, in that any given capability required by a citizen to work in a industry can be provided by the worker herself or, alternatively, they can take it from the city if available. This framework allows us to see clearly that urban phenomena occur because individuals are able to ``execute'' a recipe (e.g., a production process, a program or algorithm) if the environment is favorable, that is, if the city complements the individual. How complex a given recipe is, how capable an individual is, and how suitable the city is for executing the recipe, are the three fundamental quantities that determine the overall statistics of employment in cities, as well as other measures of urban output.\footnote{A note about terminology may be necessary here. All three measures increase in value with the number of distinct capabilities considered. Hence, they can be thought of as measures of diversity (or variety, as defined by \citep{stirling2007general}; see also \citep{van2020variety}). The more capabilities are required by an industry, or possessed by an individual or by a city, the more possibilities of recombining capabilities there are (for example, $r_c$ itself has the potential to create a constant push toward greater $r_c$ as argued in electronic supplementary material~C). This aspect of the model supports the interpretation of these quantities as measures of complexity as well (in the same vein that \citep{mcshea2010biology} characterize how biological diversity and complexity go hand-in-hand), despite the fact that diversity and complexity are typically considered to measure different aspects of a system (for a thorough review of these concepts see \citep{page20101}).}

\section{Theoretical and methodological implications of the model}\label{sec_implications}

{\subsection{Urban scaling laws}\label{sec_implications_us}}
Equipped with Eq.~(\ref{epercapita}) we can reproduce the power-law relationships of employment with population size (see \citep{YounEtAl2015Universality}). To that end, we invoke models of cultural evolution which predict a specific association between population size and the size of collective knowhow, $r_c=a+b\ln(N_c)$ (see \citep{Henrich2004Tasmanian,henrich2015secret,acerbi2017cultural}). This relation is consistent with observations in urban contexts, in which the diversity of factors and capabilities in city $c$ is approximately a logarithmic function of population size \cite{YounEtAl2015Universality}. Incorporating this relationship between $r_c$ and $N_c$ into Eq.~(\ref{epercapita}) yields $p_{i,c,f}=e^{-M_f (1-s_i)(1-a)}e^{b M_f (1-s_i)\ln(N_c)} = e^{-M_f (1-s_i)(1-a)} N_c^{b M_f (1-s_i)}$. It becomes clear why a power-law function emerges, since we are exponentiating a logarithm (see \citep{gomez2016explaining} for details about testing the predictions of this explanation). { This result indicates that more complex phenomena scale more superlinearly, explaining why more complex technological innovations and production processes tend to occur in larger and more diverse urban hubs \cite{DavisDingel2014NBER,desmet2015geography,gomez2016explaining,balland2018complex}. Relatedly, it also explains why and how different phenomena of the same kind (e.g., two sexually transmitted diseases, two types of college degrees, or two types of crimes), which presumably are driven by similar networks of social interaction, can feature very different urban scaling patterns (see \citep{Patterson2015STDs,rocha2015non,arbesman2011scaling,patterson2018scaling}). 

Although these are solid theoretical grounds on which to base the relationship between $r_c$ and $N_c$, below we will estimate directly the parameter $r_c$ from data, and show how its association with the logarithm of $N_c$ also emerges as an empirical result.}

{\subsection{Nestedness of activities across locations}\label{sec_implications_nest}
\begin{figure}[t!]
	\centering
		\includegraphics[width=0.8\textwidth, trim = 0in 0in 0in 0in]{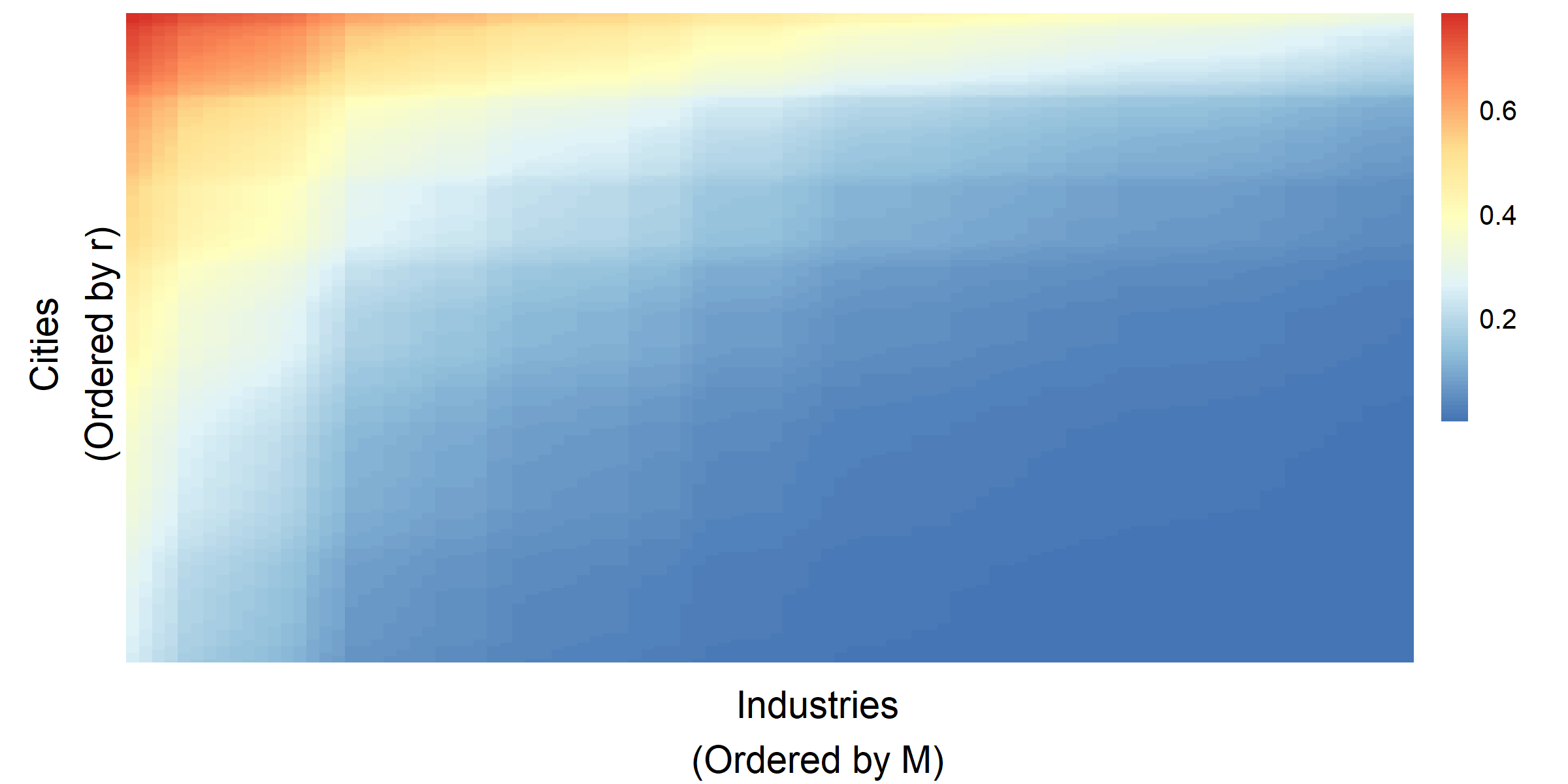}
		\caption{{\bfseries Nestedness emerges from the model.} For 100 cities and 100 industries the probability of employment is shown, with $s=0.1$ assumed constant, and with $r$ distributed uniformly between 0.7 and 0.95, and $M$ distributed uniformly between 5 and 25.}
	\label{fig_simulation_triangular}
\end{figure}}
The nested (triangular) pattern of industries across cities (see, e.g., \citep{BustosEtAl2012}) emerges naturally from the function in Eq.~(\ref{epercapita}). To show that, assume for simplicity that $s_i$ is approximately constant for all individuals in all locations. Next, let there be two industries $L$ and $H$ such that $M_L<M_H$, and two cities $l$ and $h$ such that $r_l<r_h$. According to Eq.~(\ref{epercapita}), the probability of employment is higher in the city with high collective knowhow in both industries, since $p(M_H, s, r_h) > p(M_H, s, r_l)$ and $p(M_L, s, r_h) > p(M_L, s, r_l)$. In addition, the share of employment in high complexity industries with respect to low complexity industries in diverse cities will be larger than the same ratio in less diverse cities, since $p(M_H, s, r_h)/p(M_L, s, r_h)>p(M_H, s, r_l)/p(M_L, s, r_l)$. The latter property is called ``log-supermodularity'', and reflects why diverse cities are disproportionately more competitive in complex economic activities than less diverse cities {\citep{feldman1999innovation,DavisDingel2014NBER,balland2017geography,balland2018complex,bahar2020birthplace}. This result is shown graphically in Figure~\ref{fig_simulation_triangular}, where cities and industries have been given different values of $r$ (distributed uniformly between 0.7 and 0.95) and $M$ (distributed uniformly between 5 and 25), and $p(M, s, r)$ (for constant $s=0.1$) is computed. The triangular pattern that emerges shows that probabilities of employment are nested, as empirically observed \citep{BustosEtAl2012}. According to the model, it is clear that this nested pattern can become more drastic if we assume individuals with high individual knowhow (high $s$) tend to locate themselves in cities with high $r$. This sorting of individuals is observed empirically and we discuss in the next section how it may emerge from the model.}

{\subsection{Sorting of individuals with high knowhow}\label{sec_sortinghighindividuals}
Here we show some arguments that suggest the model may reproduce the observation that} skilled individuals would tend to locate themselves in large diverse cities \cite{glaeser2010complementarity}. This is shown by noting that the probability in Eq.~(\ref{epercapita}) implies a form of assortative matching between individuals and cities, assuming an aggregate maximization of total employment across all cities subject to congestion costs (see details in electronic supplementary material~B). To show this, let there be two individuals $L$ and $H$ such that $s_L<s_H$ and two cities $l$ and $h$ such that $r_l<r_h$. The question is how will the two individuals sort themselves into these two cities, assuming there is a cost for being in the same city. Given this, we have that $p(M, s_L, r_l) + p(M, s_H, r_h) > p(M, s_L, r_h) + p(M, s_H, r_l)$, for a fixed (but large) $M_f=M$ (see electronic supplementary material~B). In words, {the most likely situation is one in which the individual with high (low) levels of knowhow will sort to places with high (low) levels of collective knowhow.} In a more complete model with prices and better specification of the costs of congestion, the phenomenon of assortative matching should ultimately be reflected in wages, as has been observed  \cite{glaeser2010complementarity,mion2009spatial}. We will explore this issue empirically by studying how our variables of economic complexity, once estimated, correlate with wages in cities across industries.

\subsection{Relationships between the three drivers of urban economic complexity}\label{sec_drivers}
{We refer to the (negative) exponent in Eq.~(\ref{epercapita}) as a ``net complexity'' (see \citep{grunwald2003kolmogorov}). This exponent is decomposable into the original three drivers of urban complexity. They are, a priori, independent quantities.\footnote{Statistically, however, we expect them to be correlated since firms with complex production processes (high $M_f$) are likely to choose to locate in diverse cities (high $r_c$), which are the places where high skill individuals (high $s_i$) sort themselves into (see electronic supplementary material~B for details).} Presently, we lack a detailed theory about the dynamical laws of {the variables $M_f$, $s_i$ and $r_c$ within our modeling framework}, and how they relate to one another. Still, it is reasonable to expect that increases in collective knowhow (through immigration that brings new capabilities and knowhow,\footnote{Models such as that in \cite{borjas2014immigration} show that when skills of immigrants are complementary to those of locals, the wages of both locals and immigrants increase.} 
 direct foreign investments that inject specific capabilities to specific industries, or by  innovation) may have a reinforcing effect that sparks and fuels a virtuous cycle: a place with a relative large enough collective knowhow will attract more people and facilitate the appearance of more complex economic activities, which will increase in turn the collective knowhow in that place.} This process will thus propel a run-away cycle of collective learning that will concentrate economic activities and wealth in large cities (see, e.g., \citep{OClery2018}). The more complex the activities, the more concentrated they will be in fewer places. Interestingly, changes in these three terms have \emph{exponential} effects in the probability of employment (more details in electronic supplementary material~C). 

The fact that the net complexity term is decomposable leads to a strategy for estimating its components {by taking logarithms twice, converting an exponential of a product, into a simple sum. This method is valid only if our model is an accurate description of how employment is created. To test the latter assumption,} in what follows, we will estimate these quantities, and demonstrate that Eq.~(\ref{epercapita}) provides important statistical improvements over alternative conceptualizations from both a modeling and a predictive power perspectives. Crucially, the resulting estimates of the city-specific driver (urban collective knowhow) is associated with measures of economic performance. In brief, we show that the ``drivers'' of urban economic complexity can be measured and have measurable consequences.

\section{Materials and methods}


{\subsection{Data}\label{sec_data_maintext}}
We use data on the estimated counts of employment, number of establishments and average wages by city-industry-year. Data were downloaded using the programming codes made available by the Bureau of Labor Statistics. See electronic supplementary material~D for additional details of the data.

\subsection{Estimating the drivers of urban complexity from data}\label{sec_estimation}
Our estimates of complexity (for industries, cities, and individuals) stand for parameters in our mathematical model. In this section, we show how to estimate these three drivers of complexity using estimated fixed effects in a regression.

Assume, for now, we know the value of the probability in Eq.~(\ref{epercapita}), ${p}_{i,c,f}$. Equating such estimate to the proposed functional form, and taking negative logarithms twice, yields

\begin{equation}
	-\ln\left(-\ln\left({p}_{i,c,f}\right)\right)=-\ln(M_f) -\ln(1-s_i) -\ln(1-r_c),
	\label{eq_fefullmodel}
\end{equation}
Equation~(\ref{eq_fefullmodel}) shows how net complexity is decomposed linearly into its main components. To estimate the value of the economic complexity variables in our model, Eq.~(\ref{eq_fefullmodel}) implies we can regress $-\ln\left(-\ln\left({p}_{i,c,f}\right)\right)$ against three \emph{additive} fixed-level effects corresponding to the activity, the individual, and the city. 

The regression method we propose, however, has some limitations in practice. As it is typically the case, microdata at the individual level is often difficult to obtain, which makes ${p}_{i,c,f}$ difficult to estimate. When the only information available is aggregate counts of employment per industry and city, we cannot estimate the probability on the left-hand side of Eq.~(\ref{eq_fefullmodel}) at the level of each individual, and therefore we cannot include an individual-fixed effect on the right-hand side. 
 
Fortunately, the model is simple enough that we can address this limitation by substituting $s_i$ in Eq.~(\ref{epercapita}) with the \emph{average} individual knowhow in city $c$, $\bar{s}_c \approx \sum_{i\in c}s_i/N_c$. This substitution implies that $\Pr\{X_{c,f}=1\}=\exp\left(-M_f(1-\bar{s}_c)(1-r_c)\right)$ (see electronic supplementary material~E). This expression represents the probability that \emph{any} random individual in city $c$ (as opposed to a specific one) is employed in activity $f$. We estimate $p_{c,f}=\Pr\{X_{c,f}=1\}$ through the share of employment $\widehat{p}_{c,f} = y_{c,f}=Y_{c,f}/N_c$, where $Y_{c,f}$ is the employment count in industry $f$ in city $c$ and $N_c$ is total population size.\footnote{The estimate of $p_{c,f}$ could in principle incorporate a Bayesian prior using pseudocounts, $\widehat{p}_{c,f} = (Y_{c,f}+\alpha_{c,f})/(N_c + \sum_{f}\alpha_{c,f})$ in order to handle the case when $Y_{c,f}=0$. See \citep{van2020information}.} 
 The regression equation that will allow us to estimate the complexity of industries and the collective knowhow of cities (but not the individual knowhow of workers) becomes
\begin{align}
	-\ln\left(-\ln\left(y_{c,f}\right)\right)&=\delta_f + \gamma_c + \varepsilon_{c,f},
\label{e2LOGLOGpercapita}
\end{align}
where $\varepsilon_{c,f}$ is the error term, $\delta_f$ stands for $-\ln(M_f)$ and $\gamma_c$ for $-\ln((1-\bar{s}_c)(1-r_c))$.\footnote{{If one has an estimate of $\widehat{p}_{c,f}$ for all possible combinations of $c$ and $f$ in the data (i.e., there are no missing values in the matrix of places and activities), the method of estimation is even simpler. Applying a singular value decomposition (SVD), or principal component analysis (PCA), to $\log(\widehat{p}_{c,f})= -M_f(1-\bar{s}_c)(1-r_c)$, will yield a single pair of vectors, one vector of principal components and another of principal loadings, corresponding to the vector of $(1-\bar{s}_c)(1-r_c)$ and $M_f$, respectively. Since missing values in data are more the rule than the exception, the method of fixed effects proposed in this paper is likely to have greater applicability than the application of SVD.}} We note that the city fixed effect is a city-specific variable that includes the interaction between the suitability of the urban environment and the average capacity of citizens. While we are unable to learn much about individual-level knowhow from our data, the inclusion of a proxy allows us to learn about the effect that the other two drivers have on measures of urban performance (Section~\ref{sec_linkingdrivers}).

\subsection{Evaluation against competing models} \label{sec_competingmodels}
We evaluate the predictive power of model \eqref{e2LOGLOGpercapita} with respect to four alternative models using holdout data. Two of the models differ from our model only in terms of the functional transformation of the dependent variable; that is, rather than $-\ln(-\ln(y_{c,f,t}))$, the left-hand side is given by $y_{c, f, t}$ (namely Model $1.1$) and $\ln(y_{c, f, t})$ (Model $1.2$), respectively. As shown in electronic supplementary material~F, Model $1.2$, for example, is derived from a world in which capabilities are not complementary (as we have assumed) but substitutable. As a consequence, differences in predictive performance between model \eqref{e2LOGLOGpercapita} and these two other models will inform us about the importance of the specific functional form predicted by the theory and its underlying assumptions. 

The other two alternative models differ from \eqref{e2LOGLOGpercapita} in the formulation of the right-hand side of the model. One model (namely Model $2.1$) will be the standard urban scaling model where it is assumed that the scaling exponent is the \emph{same} for all phenomena, as suggested by network-based explanations \cite{Bettencourt2013}. The last alternative model (Model $2.2$) is an unconstrained version of the standard scaling model, where it is assumed that both the baseline prevalence and the scaling exponent differ, in principle, for each industry $f$ (see electronic supplementary material~F). Differences in performance between model \eqref{e2LOGLOGpercapita} and these last two models will inform us about the validity of adding degrees of freedom to explain employment patterns across cities and industries.  

To compare these models based on ``out-of-sample'' prediction performance, for each year, we split the data into training and testing (or validation) sets. After the parameters of the models are fitted using the training data, they are then compared by how accurately they predict the dependent variable on the test set. The predictions were evaluated using the root mean squared error (\emph{RMSE}) and the mean absolute error (\emph{MAE}). The train and test random splits were repeated $100$ different times of the data (bootstrapping cross-validation). See electronic supplementary material~F for more details.

\subsection{Linking the drivers with urban economic performance}\label{sec_linkingdrivers}
There is a tension between individuals and firms trying to reap the benefits of diversity in cities, and individuals and firms trying to avoid each other to reduce congestion costs from living and operating in the same city. As we argued in Section~\ref{sec_sortinghighindividuals}, however, we expect skilled individuals and complex industries to locate themselves in large cities, paying higher wages to compensate for the costs of congestion. Thus, in the empirical analysis that we propose in this section, we are interested in analyzing whether our variables of interest (industry complexity and collective knowhow) are \emph{positively} associated with measures of economic performance of firms and workers. We will measure performance with average wages and average size of establishments at the level of city-industry combinations.\footnote{Here, we face one challenge regarding what the data measure. The Bureau of Labor Statistics provides data about establishments, not firms, at the level of city-industry combinations. Hence, a large firm may be composed of many small establishments, and these may affect the validity of our regression model Eq.~(\ref{eq_wageregmodel2}).}

{Given that firms and workers in larger (more populous) cities are more productive \cite{Sveikauskas1975,glaeser2001cities,RosenthalStrange2004,DurantonPuga2004micro,MeloGrahamNoland2009meta,CombesDurantonGobillon2008,Combes2012productivity,Polese2005reappraisal,Polese2013,andersson2014sources,Behrens2014productive}, we also included in our regressions the logarithm of city population size as a control variable. In addition, we will include two measures of city and industry complexity that have been proposed before in the literature on economic complexity. These measures are known as the Economic Complexity Index (ECI) and the Product Complexity Index (PCI) \cite{HidalgoHausmann2009}, here applied to cities and industries (instead of countries and export products).\footnote{These indices are based on a spectral clustering method \cite{shi2000normalized,newman2006finding,von2007tutorial}. As such, the method provides one vector of ratings for cities and another for industries, and these ratings cluster cities and industries according to their pairwise similarities deduced from the presence/absence matrix of industries across cities.} These indices are not statistical estimates of knowhow or complexity. Instead, they are a mathematical construction which, given their empirical high correlation with measures of economic growth \citep{fritz2021economic}, are assumed to rank cities and industries by their underlying number of capabilities (see \citep{mealy2019interpreting} for a discussion, and \citep{schetter2019structural} for  analytic justifications of this method). %

We will analyze these associations through linear regressions of the following form: 
\begin{align}
	\ln(z_{c,f,t})=\beta_0 + \beta_1\ln(\bar{s}^{\text{proxy}}_{c,t}) - \beta_2\widehat{\delta_{f,t}} + \beta_3\widehat{\gamma_{c,t}} + {\mathbf X}_{\text{controls}}{\bm \beta}_{\text{controls}} + \varepsilon_{c,f,t}.
	\label{eq_wageregmodel2}
\end{align}}
where $z_{c,f,t}$ stands for either wages or size of establishments, and where we have made explicit the time dimension $t$ representing years. We will carry out different specifications of regression model (\ref{eq_wageregmodel2}) for different combinations of the explanatory variables.\footnote{ For simplicity of presentation, we have not included diagnoses of spatial autocorrelations. However, researchers seeking to explain urban outcomes should take these into consideration, since cities are not independent of one another.} In the main text we will show the results for one specific year, but in electronic supplementary material~L we show the results for all years, conducting one regression for each year separately to control for changes in nominal prices. 

{As a consequence of lacking individual-level microdata, and as explained in Section~\ref{sec_estimation}, we do not have an estimate of individual knowhow that we can include in Eq~.(\ref{eq_wageregmodel2}). However, we reduce the potential effects of omitted variable bias by including a quantity that acts as a proxy for individual knowhow in Eq~.(\ref{eq_wageregmodel2}), $\ln(\bar{s}^{\text{proxy}}_{c,t})$. For this, we consider educational attainment. More years of educational attainment are usually a measure of specialization, but they {can also be} an indication of a person's competency to perform several productive tasks, as our model assumes. Whether years of schooling is a good proxy for individual knowhow, however, remains an empirical question (see, e.g., \citep{bacolod2010elements}). Acknowledging these caveats, we will proxy $\bar{s}_{c,t}$ using the average levels of schooling in the city $c$ in year $t$ (see electronic supplementary material~E for details). {Thus, statements and hypotheses about how, based on our model, this driver affects wages and establishment sizes should be taken with caution.}}

{Finally, we note that we use as independent variables in Eq.~(\ref{eq_wageregmodel2}) the estimates of industry complexity and collective knowhow, $-\widehat{\delta_{f,t}}$ and $\widehat{\gamma_{c,t}} $, coming from the results of Section~\ref{sec_estimation}. Moreover, we note the use of a negative for $\widehat{\delta_{f,t}}$ and a positive one for $\widehat{\gamma_{c,t}}$ such that their interpretation is consistent with measures of complexity and knowhow.}

\section{Results}

\subsection{Comparing prediction power of models}

\begin{figure}[t!]
	\centering
\includegraphics[width=0.95\textwidth, trim = 0in 0in 0in 0in]{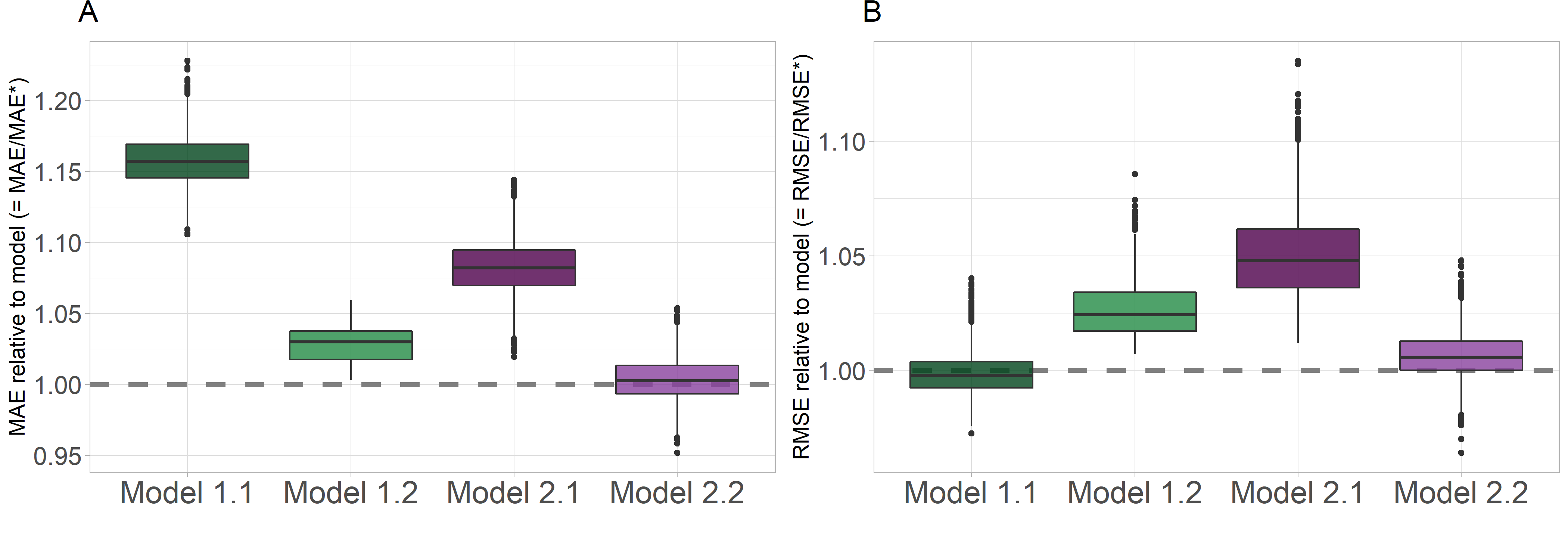}
		\caption{{\bfseries Evaluation against competing models.} Comparison of out-of-sample predictions from all models using $100$ random cross-validation train/test splits. {\bfseries A.} $y$-axis represents the ratio between the mean absolute error ($\mathrm{MAE}$) of each alternative model divided by the $\mathrm{MAE}^*$ of model \eqref{e2LOGLOGpercapita}. {\bfseries B.} $y$-axis represents the ratio between the root mean square error ($\mathrm{RMSE}$) of each alternative model divided by the $\mathrm{RMSE}^*$ of model. {For both {\bfseries A} and {\bfseries B}, values larger than one represent worse out-of-sample predictions as compared to our model.}} 
	\label{Fig_densities_threemodels}
\end{figure}

Figure~\ref{Fig_densities_threemodels} presents a comparison of our model \eqref{e2LOGLOGpercapita} against four alternative models, as specified in Section~\ref{sec_competingmodels}. To make the comparisons clearer, in the figure, we have divided the \emph{MAE} score (and \emph{RMSE}) of each alternative model by the \emph{MAE}$^*$ (\emph{RMSE}$^*$) of our model in each run of the bootstrapping cross-validations. Models with worse predictive performance than ours will be above the dashed line.

Results show that our model has superior performance in terms of ${MAE}$ except for the unconstrained scaling model (Model $2.2$, where baseline prevalence and the scaling exponent can differ for each industry $f$). The comparable performance of Model $2.2$ is supportive of the ideas and results put forth by \citep{gomez2016explaining}. Our model has also superior performance in terms of ${RMSE}$ with respect to all models except for the first alternative model (Model $1.1$, where the dependent variable is not logged). Interestingly, this indicates that city and industry fixed effects directly fitted on per capita employment provide an alternative good fit to the data, but only when there are no extreme values ($RMSE$ is more sensitive to outliers).

\subsection{Relation of collective knowhow to population size and industrial diversity}

Having estimated $\gamma_{c,t}$ and $\delta_{c,t}$, as described in Section~2.1, we present a descriptive understanding of how much these estimates conform to our proposed notions of collective knowhow and industry complexity. 
\begin{figure}[t!]
	\centering
		\includegraphics[width=0.9\textwidth, trim = 0in 0in 0in 0in]{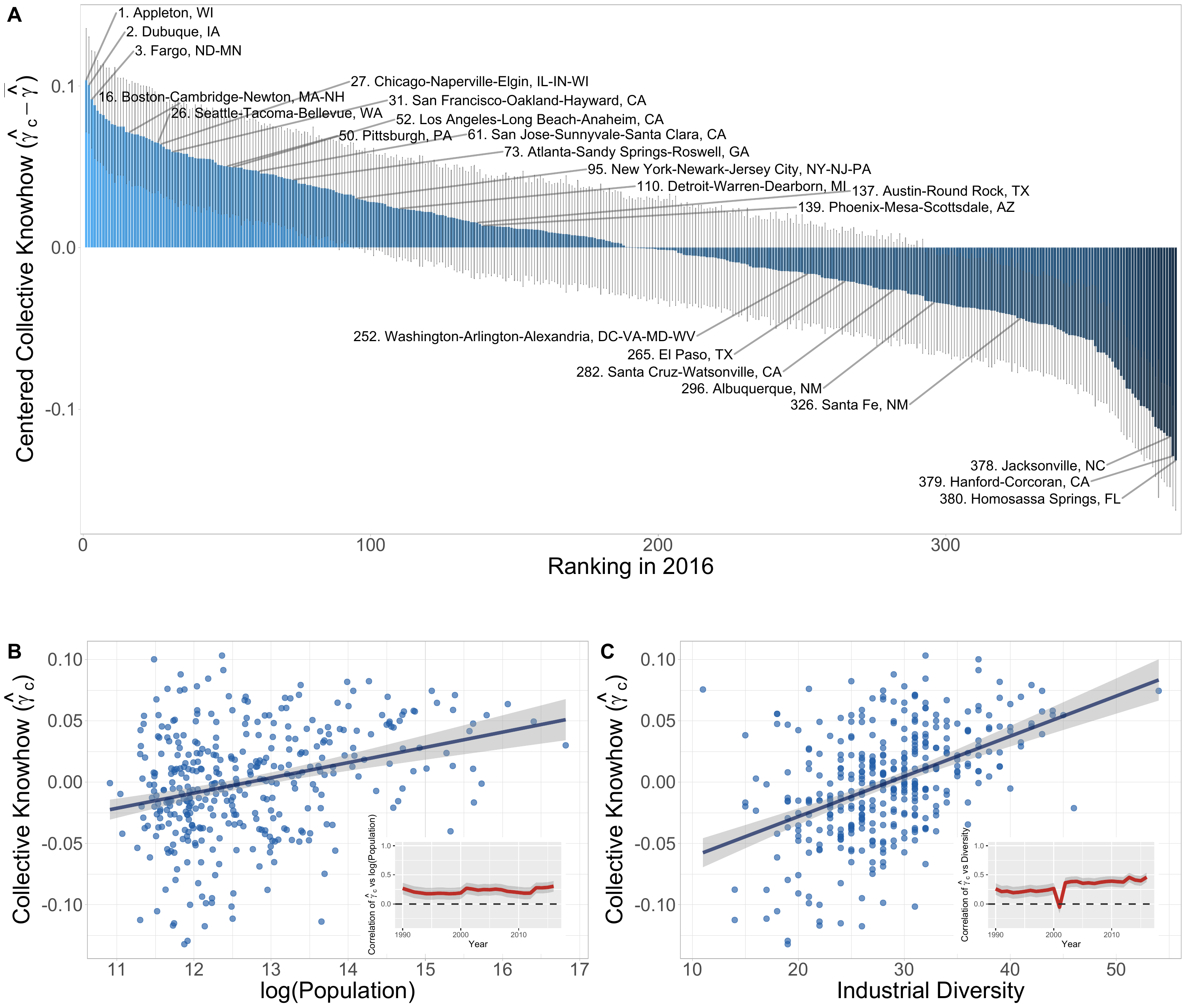}
		\caption{{\bfseries Estimated collective knowhow across US cities.} {\bfseries A.} Bar plot and the ranking of cities according to their centered scores of collective knowhow in 2016. The plot shows $\pm$ one standard errors coming from the regression estimation for every city. {\bfseries B.} Association with (log) population size in 2016 {(pearson $\rho=0.30$, $R^2=0.09$)}. The inset shows the pearson correlation across years. {\bfseries C.} Association with industrial diversity in 2016  {(pearson $\rho=0.46$, $R^2=0.21$)}. The inset shows the pearson correlation across years.} 
	\label{Fig_associations_ckh}
\end{figure}

Figure~\ref{Fig_associations_ckh}A shows that most large cities have scores of collective knowhow that are above the average. Among the large cities, Boston has the largest score. One exception is the metropolitan area of Washington DC, which falls below the average. Interestingly, New York City is closer to cities like Detroit, Austin and Phoenix, than to cities like Chicago, Los Angeles or Boston. The estimates of collective knowhow, however, have large standard errors (shown as gray segments in each bar) due to the fact that the estimation of the fixed effect relies on few industries present per city (thirty industries per city on average).

Figure~\ref{Fig_associations_ckh}B and C show how the scores correlate with population size and industrial diversity, respectively. Here, we define industrial diversity by the number of industries in each city that have location quotients larger than one. That is, if $LQ_{c,f}=(Y_{c,f}/\sum_{f} Y_{c,f})/(\sum_{c} Y_{c,f}/\sum_{c,f} Y_{c,f})$, then diversity is $d_c = \sum_f \mathbf{1}_{\left(LQ_{c,f}>1\right)}$ (e.g., see \cite{HidalgoHausmann2009}). Our estimate of $\widehat{\delta_{c,t}}$ increases with both population size and industrial diversity, supporting the assumption that it fits the pattern of a measure of collective knowhow. The association with diversity, interestingly, has been increasing with time (see inset of Figure~\ref{Fig_associations_ckh}C), which may suggest the fact that the process of diversification is driven by larger bodies of collective knowhow.

\subsection{Relation of industry complexity to occupational diversity, and geographical ubiquity}

\begin{figure}[t!]
	\centering
		\includegraphics[width=0.9\textwidth, trim = 0in 0in 0in 0in]{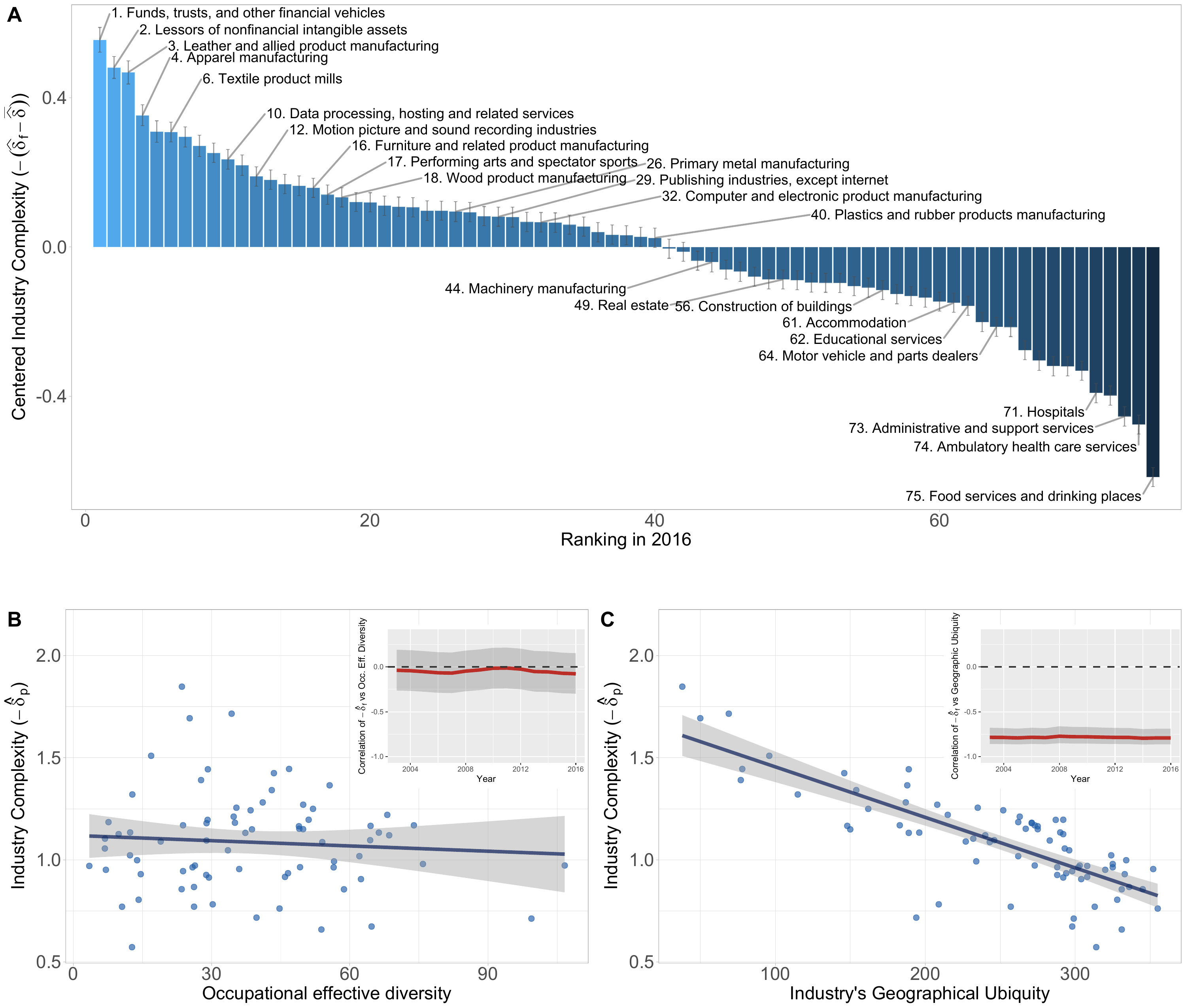}
		\caption{{\bfseries Estimated industry complexity across 3-digit NAICS industries.} {\bfseries A.} Bar plot and the ranking of industries according to their centered scores of industry complexity in 2016. {We note that we show the negative sign next to $\widehat{\delta}_f$ so that higher levels are interpreted as higher complexities.} The plot shows $\pm$ one standard errors coming from the regression estimation for every industry. {\bfseries B.} Association with the effective number of occupations per industry in 2016 {(pearson $\rho=-0.08$, $R^2=0.01$)}. The inset shows the pearson correlation across years. {\bfseries C.} Association with the geographical ubiquity per industry in 2016 {(pearson $\rho=-0.79$, $R^2=0.62$)}. The inset shows the pearson correlation across years.} 
	\label{Fig_associations_inds}
\end{figure}

Next, we assess whether our estimate of the complexity of industries is consistent with our notion of ``difficulty''. Thus, we compare our estimates with a measure of occupational diversity on the one hand, and with a measure of geographical concentration on the other. We define ``occupational effective diversity'' based on the shares of employment across occupations per industry, computing it by taking the exponential of Shannon's entropy. That is, if $p_{f,o}=E_{f,o}/\sum_{o} E_{f,o}$ is the U.S. national share of employment of occupation $o$ in industry $f$, then the occupational effective diversity is $d_f=\exp\left\{-\sum_o p_{f,o}\ln(p_{f,o})\right\}$ (see \citep{jost2006entropy,jost2007partitioning,van2019diversity}, and electronic supplementary material~D for the source of occupation-industry data). The measure of geographical concentration, ``ubiquity'', is analogous to above's measure of industrial diversity for cities. That is, $u_f = \sum_c \mathbf{1}_{\left(LQ_{c,f}>1\right)}$ (e.g., see \cite{HidalgoHausmann2009}). 

Figure~\ref{Fig_associations_inds}A shows that many financial and manufacturing-related industries have scores that are above the average. Interestingly, ``Performing arts and spectator sports'' appears in ranking 17, above industries like ``Computer and electronics'' or ``Plastics and rubber products'' manufacturing. The least complex industries, not unsurprisingly, are some common service industries, with ``Food services and drinking places'' at the bottom of all. 

Figure~\ref{Fig_associations_inds}B and C show how the scores correlate with the measure of occupational diversity of industries, and with the measure of geographical ubiquity, respectively. Somewhat unexpectedly, the measure of complexity does not correlate with the effective number of occupations that are typically employed by the industry. This result may be a consequence of our model not taking into account the other economic forces at play, but it may also suggest that occupations are not the fundamental units of knowhow. Conversely, the fact that the most complex industries occur in few metropolitan areas indicates that these industries are indeed dependent on the right urban context.

\subsection{Relationship between the drivers and the establishment sizes and wages}

\begin{figure}[t!]
	\centering
		\includegraphics[width=0.9\textwidth, trim = 0in 0in 0in 0in]{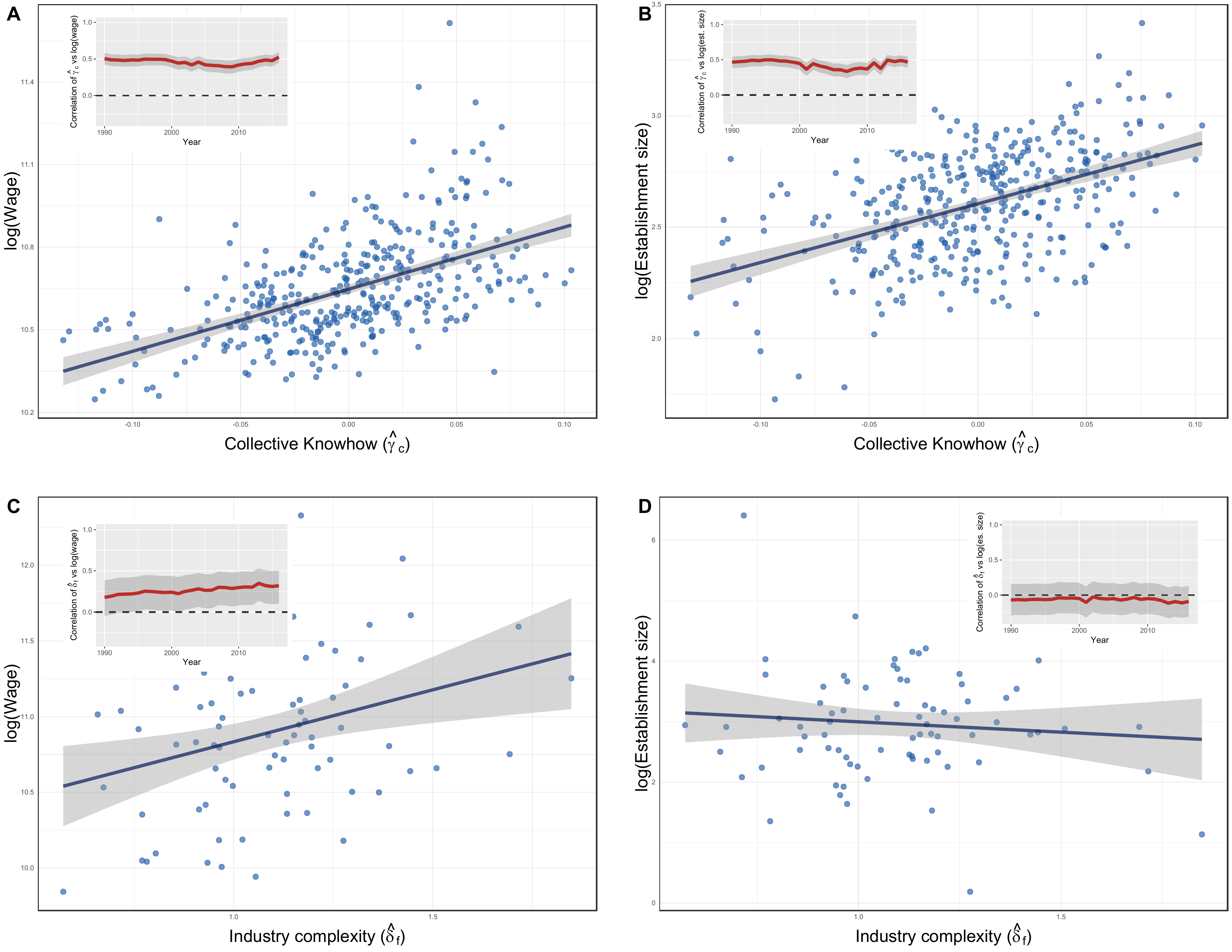}
		\caption{{\bfseries A.} Association of collective knowhow with (log) average wages at the level of cities {(pearson $\rho=0.53$, $R^2=0.28$)}. {\bfseries B.} Association of collective knowhow with (log) average establishment size at the level of cities {(pearson $\rho=0.47$, $R^2=0.22$)}. {\bfseries C.} Association of industry complexity with (log) average wages at the level of industries {(pearson $\rho=0.32$, $R^2=0.10$)}. {\bfseries C.} Association of industry complexity with (log) average establishment size at the level of industries {(pearson $\rho=-0.09$, $R^2=0.01$)}. All scatter plots are for 2016 data, and all insets show the pearson correlation across all years.} 
	\label{Fig_correlations_ckh}
\end{figure}

{Here we show the associations of our estimates with wages and establishment sizes at two levels of aggregation: coarse (each observation is either a city or an industry) and fine-grained (each observation is a city-industry combination).}

We start this section by showing the correlations when observations are aggregated at the city level and at the industry level, separately. These associations are shown in Figure~\ref{Fig_correlations_ckh}. In panels A and B each point is a city, and the figures indicate that the association between average wages and average establishment size with collective knowhow is very strong. The association is stable across years, as shown in the respective insets. In panels C and D each point is an industry, and the figures show that the association of wages and establishment sizes at the aggregate level of industries, however, is less strongly associated with our estimate of industry complexity.

{Next, we investigate the associations at a fine-grained level of aggregation in which each observation is a city-industry combination, as proposed in Section~\ref{sec_linkingdrivers} in Eq.~(\ref{eq_wageregmodel2}). At this level of disaggregation, there were 19,490 different city-industry combinations in our data in 2016 (out of $381\times 78 = 29,718$ possible combinations). The median number of employees per observation was 772 and the median number of establishments was 51. The question is whether our complexity variables can account for some of the variation in the average wages of workers and the average size of establishments across all these observations, and whether they remain statistically significant after adding other competing controls.} 

Tables~\ref{tab_firmsize_regs} and \ref{tab_wage_regs} present the results of seven regression models applied to the year $t=2016$, where the dependent variables are logarithm of establishment sizes and wages, respectively. {The key observation is that the increase in the level of disaggregation makes the problem of predicting wages and establishment sizes more challenging (reflected in the low $R^2$ values), and yet, industry complexity and the city knowhow remain statistically significant ($p<0.05$) in the presence of other controls like population size, PCI and ECI.}

\begin{table}[!t] \centering 
  \caption{{\bfseries Associations with average size of establishments.} Linear regressions for (log) average firm size at the city-industry level as a function of city (log) population size, (log) of average years of schooling, (log) of the inherent industry complexity, (log) of the collective knowhow of city, controlling for economic and product complexity indices. Regression table shows only the year 2016.} 
  \label{tab_firmsize_regs} 
\scriptsize
\begin{tabular}{@{\extracolsep{-4pt}}lccccccc} 
\\[-1.8ex]\hline 
\hline \\[-1.8ex] 
 & \multicolumn{7}{c}{\textit{Dependent variable:}} \\ 
\cline{2-8} 
\\[-1.8ex] & \multicolumn{7}{c}{log(Ave. Establishment Size)} \\ 
\\[-1.8ex] & (1) & (2) & (3) & (4) & (5) & (6) & (7)\\ 
\hline \\[-1.8ex] 
 log(City Population) & 0.152$^{***}$ &  &  &  &  & 0.140$^{***}$ & 0.131$^{***}$ \\ 
  & t = 24.865 &  &  &  &  & t = 22.224 & t = 17.029 \\ 
  & & & & & & & \\ 
 log(Ave. Yrs. Schooling) &  & 0.397$^{***}$ &  &  & $-$0.563$^{***}$ & $-$0.842$^{***}$ & $-$0.932$^{***}$ \\ 
  &  & t = 4.630 &  &  & t = $-$6.096 & t = $-$9.144 & t = $-$9.336 \\ 
  & & & & & & & \\ 
 log(Industry Complexity) &  &  & $-$0.852$^{***}$ &  & $-$0.888$^{***}$ & $-$0.946$^{***}$ & $-$0.770$^{***}$ \\ 
  &  &  & t = $-$24.610 &  & t = $-$26.057 & t = $-$28.018 & t = $-$21.118 \\ 
  & & & & & & & \\ 
 log(City Collective Knowhow) &  &  &  & 3.750$^{***}$ & 4.387$^{***}$ & 3.539$^{***}$ & 3.324$^{***}$ \\ 
  &  &  &  & t = 23.947 & t = 25.639 & t = 20.429 & t = 17.861 \\ 
  & & & & & & & \\ 
 PCI &  &  &  &  &  &  & $-$0.025$^{***}$ \\ 
  &  &  &  &  &  &  & t = $-$12.654 \\ 
  & & & & & & & \\ 
 ECI &  &  &  &  &  &  & 0.031$^{***}$ \\ 
  &  &  &  &  &  &  & t = 3.005 \\ 
  & & & & & & & \\ 
 Constant & 0.691$^{***}$ & 2.065$^{***}$ & 2.644$^{***}$ & 2.630$^{***}$ & 3.445$^{***}$ & 2.055$^{***}$ & 2.317$^{***}$ \\ 
  & t = 8.830 & t = 16.919 & t = 382.560 & t = 381.521 & t = 26.203 & t = 14.257 & t = 12.710 \\ 
  & & & & & & & \\ 
\hline \\[-1.8ex] 
Observations & 19,490 & 19,490 & 19,490 & 19,490 & 19,490 & 19,490 & 19,490 \\ 
R$^{2}$ & 0.031 & 0.001 & 0.030 & 0.029 & 0.063 & 0.086 & 0.094 \\ 
Adjusted R$^{2}$ & 0.031 & 0.001 & 0.030 & 0.029 & 0.063 & 0.086 & 0.094 \\ 
\hline 
\hline \\[-1.8ex] 
\textit{Note:}  & \multicolumn{7}{r}{$^{*}$p$<$0.05; $^{**}$p$<$0.01; $^{***}$p$<$0.005} \\ 
\end{tabular} 
\end{table}

In Table~\ref{tab_firmsize_regs}, the coefficient for population size is fairly constant across specifications, indicating that in the cross-section of U.S. metropolitan areas, a $1\%$ increase in population size is, on average, associated with $0.13-0.15\%$ increases in average establishment size. Average years of schooling is positively associated with establishment size, but when we control for the rest of variables, it switches sign and becomes negatively associated, indicating that a $1\%$ increase in the average years of schooling of individuals in a city is associated with a $0.6-0.9\%$ \emph{reduction} of establishment size. Such result, although somewhat unintuitive, is consistent with our theory. However, we remind the reader that average years of schooling is not a direct estimate of individual knowhow, since we did not use our methodology to estimate that parameter of our model, and serves here a statistical role to reduce the effect of omitted variables. 

The association between firm size and the inherent complexity of industries, however, is unexpected. We find that a $1\%$ increase in the number of ``ingredients'' of an industry is associated with approximately a $0.8\%$ reduction in the size of establishments. Note that this negative relationship also holds for the Product Complexity Index. At face value, these results may be indications of an inconsistency in our model, or an indication that there may be unaccounted sources of error or omitted variables. Together with Figure~\ref{Fig_associations_inds}B, these results suggest that investigating the determinants of establishment size across different industries, and their connection with economic complexity, is an interesting direction for future work. 

Finally, the relationship with collective knowhow is as expected: establishments located in cities with high levels of collective knowhow are larger in size. This positive relationship is maintained even after including the Economic Complexity Index.


\begin{table}[!t] \centering 
  \caption{{\bfseries Associations with average wages.} Linear regressions for (log) average wages at the city-industry level as a function of city (log) population size, (log) of average years of schooling, (log) of the inherent industry complexity, (log) of the collective knowhow of city, controlling for economic and product complexity indices. Regression table shows only the year 2016.} 
  \label{tab_wage_regs} 
\scriptsize
\begin{tabular}{@{\extracolsep{-6pt}}lccccccc} 
\\[-1.8ex]\hline 
\hline \\[-1.8ex] 
 & \multicolumn{7}{c}{\textit{Dependent variable:}} \\ 
\cline{2-8} 
\\[-1.8ex] & \multicolumn{7}{c}{log(Ave. Wage)} \\ 
\\[-1.8ex] & (1) & (2) & (3) & (4) & (5) & (6) & (7)\\ 
\hline \\[-1.8ex] 
 log(City Population) & 0.105$^{***}$ &  &  &  &  & 0.088$^{***}$ & 0.071$^{***}$ \\ 
  & t = 34.241 &  &  &  &  & t = 27.370 & t = 18.535 \\ 
  & & & & & & & \\ 
 log(Ave. Yrs. Schooling) &  & 0.812$^{***}$ &  &  & 0.547$^{***}$ & 0.371$^{***}$ & 0.231$^{***}$ \\ 
  &  & t = 18.700 &  &  & t = 11.473 & t = 7.852 & t = 4.650 \\ 
  & & & & & & & \\ 
 log(Industry Complexity) &  &  & 0.359$^{***}$ &  & 0.341$^{***}$ & 0.304$^{***}$ & 0.043$^{*}$ \\ 
  &  &  & t = 20.188 &  & t = 19.381 & t = 17.583 & t = 2.389 \\ 
  & & & & & & & \\ 
 log(City Collective Knowhow) &  &  &  & 1.508$^{***}$ & 0.995$^{***}$ & 0.460$^{***}$ & 0.294$^{***}$ \\ 
  &  &  &  & t = 18.769 & t = 11.280 & t = 5.183 & t = 3.174 \\ 
  & & & & & & & \\ 
 PCI &  &  &  &  &  &  & 0.037$^{***}$ \\ 
  &  &  &  &  &  &  & t = 36.753 \\ 
  & & & & & & & \\ 
 ECI &  &  &  &  &  &  & 0.026$^{***}$ \\ 
  &  &  &  &  &  &  & t = 5.028 \\ 
  & & & & & & & \\ 
 Constant & 9.243$^{***}$ & 9.433$^{***}$ & 10.581$^{***}$ & 10.587$^{***}$ & 9.804$^{***}$ & 8.927$^{***}$ & 9.317$^{***}$ \\ 
  & t = 234.625 & t = 152.728 & t = 2,986.174 & t = 2,993.696 & t = 144.627 & t = 120.891 & t = 102.799 \\ 
  & & & & & & & \\ 
\hline \\[-1.8ex] 
Observations & 19,490 & 19,490 & 19,490 & 19,490 & 19,490 & 19,490 & 19,490 \\ 
R$^{2}$ & 0.057 & 0.018 & 0.020 & 0.018 & 0.043 & 0.078 & 0.140 \\ 
Adjusted R$^{2}$ & 0.057 & 0.018 & 0.020 & 0.018 & 0.043 & 0.078 & 0.140 \\ 
\hline 
\hline \\[-1.8ex] 
\textit{Note:}  & \multicolumn{7}{r}{$^{*}$p$<$0.05; $^{**}$p$<$0.01; $^{***}$p$<$0.005} \\ 
\end{tabular} 
\end{table} 


In Table~\ref{tab_wage_regs} we observe very similar patterns in wages. The coefficient for population size is again relatively stable across specifications, with a $1\%$ increase in population size associated with a $0.07-0.10\%$ increase in average wages in the city-industry cell. Average years of schooling is positively associated with wages, with a $1\%$ increase leading to a $0.2-0.8\%$ increase in wages. Also consistent with our expectations is that a $1\%$ change in the inherent complexity of an industry is associated with approximately a $0.3\%$ positive change in wages. Notice, however, that this positive association loses some statistical significance if we control for the PCI and ECI. Finally, there is a positive relationship between average wages and collective knowhow, maintained even after including the ECI. However, this relationship also loses some statistical significance after we control for city population size.


The results shown in Tables~\ref{tab_firmsize_regs} and \ref{tab_wage_regs} are unchanged across the years (see electronic supplementary material~L, Figure~14). Interestingly, the time evolution of these coefficients reveal that wages and industry complexity have become more tightly associated over the years.

\section{Discussion and concluding remarks}
{
Work in economic complexity and evolutionary economics has proposed that economic development is the process of accumulating an ever increasing variety of capabilities \citep{NelsonWinter1982,HidalgoHausmann2009,boschma2017relatedness,HausmannHidalgo2011,hidalgo2021economic}, as opposed to a process of increasing the intensity of a few factors of production. However, since productive capabilities are often embedded in people as tacit knowledge \citep{johnson2002all,kogut1992knowledge,Neffke2013SkillRelatedness,coscia2020knowledge}, and are therefore not always observable or identifiable, \citet{HidalgoHausmann2009} first introduced a set of indicators to infer the number of capabilities present in a place (and required by firms in different economic activities) by analyzing directly the high-dimensional data about the collection of products that places are able to produce. They named these indicators the Economic Complexity Index (ECI) and the Product Complexity Index (PCI). Soon after, additional research followed suit and developed other algorithms to use such high-dimensional data to compute complexity metrics (e.g., \citep{TacchellaCristelliCaldarelliEtAl2012SciRep,sciarra2020reconciling,brummitt2020machine}). 

While the idea of economic performance as a matter of capability accumulation is well rooted in theories of cultural and social evolution \citep{henrich2015secret}, and it is consistent with the observation that differences in economic performance across time and space seem to arise because some societies are collectively more productive than others \citep{mokyr2002gifts,henrich2015secret,hausmann2016economic}, the economic complexity indices proposed to this date rest on less solid ground, and their use remains therefore contested \citep{mariani2015measuring,Mealy2017,morrison2017economic}. So far, there is no theory that demonstrates that the mathematical manipulations that underlie these algorithms actually quantify anything like `number of capabilities'. \citet{schetter2019structural} has shown that ECI (or a measure closely related to it) may rank places according to some underlying productivity (conditional on some assumptions about the structure of the data), but others have shown that ECI is better interpreted as an index to quantify how much a place is specialized on a few economic activities \citep{gomez2018methods,mealy2019interpreting,van2021correspondence}. If we believe the economic performance of places is related to the number of capabilities present in them, then we must devise reliable ways of estimating them. As indices of economic complexity become more commonly used in the policy sphere \citep{balland2019smart,escobari2019growing,zaccaria2018integrating}, we must ensure to have sensible theories, robust methodologies, and reliable interpretations of metrics that actually quantify capabilities. In this paper, we have proposed a mathematical model combined with a data-driven approach to address these issues.

}

Our contributions consist of three main results. First, using the economic complexity framework, we derived a probabilistic model that explains the patterns of industrial employment in cities as a process of capability recombination. Second, based on the model, we proposed a simple statistical method to estimate the economic complexity variables that emerge from this model, such as the complexity of industries and the collective knowhow of cities. And third, using estimates of industry complexity and city knowhow, we showed that these variables have strong statistical associations with measures of economic performance such as average wages and average establishment sizes.

In the model, individual knowhow, industry complexity, and urban collective knowhow interact as a product in an exponential function, which makes them tightly intertwined. We showed how, from this mathematical interaction, the triangular pattern in the matrix of places and economic activities, {which has been widely observed and investigated in the literature on economic complexity (see, e.g., \cite{HidalgoHausmann2009,HausmannHidalgo2011,BustosEtAl2012,cristelli2015heterogeneous,mealy2019interpreting}) emerges (Figure~\ref{fig_simulation_triangular}). We also showed how this interaction may be interpreted as a complementarity that would tend to sort complex industries and skilled individuals together into diverse cities. This is a topic that has been widely investigated in the urban economics literature (see, e.g., \citep{mion2009spatial,glaeser2001cities}) where it has been noted that this sorting occurs as a result of the better possibilities of matching employers with employees in large cities (see, e.g., \citep{DurantonPuga2004micro}), and thus it is not surprising to find such effect in our model as well. While further exploration is needed to establish connections with other models of cities, we nevertheless} presented supporting evidence that the functional form predicted by the model is statistically superior to some alternative models. The main lesson from our model is that {a simple probabilistic `production-recipes' approach of recombination of inputs in cities, with little assumptions about the nature of those inputs or the presence of market forces, can go a long way into providing firm footing for considering economic complexity quantities as fundamental (as opposed to \emph{ad-hoc} summary statistics of high-dimensional data), for establishing a method for estimating them, and for providing reliable interpretations of these quantities as meaningful economic indicators.
}

A more practical aspect of our results was the method to estimate the parameters of the model from data, which can be extended to apply when individual data are available. In particular, we showed how, conditional on some manipulations of the data, complexity metrics can be estimated as fixed effects in a regression model. This method is likely to work best with social security data that tracks individuals and their job trajectories across places and industries. Here, however, we focused instead on analyzing estimates of industry complexity and collective knowhow.

We have shown that our estimates of the knowhow of cities correlate with intuitive measures of urban complexity like population size and industrial diversity, and they also strongly correlate with measures of urban output, like average wages and firm sizes. Crucially, however, our estimates resulted in rankings of cities by collective knowhow that were somewhat surprising, in the sense that large cities like New York or San Francisco MSAs did not rank particularly high. One explanation is that our estimates of urban collective knowhow have a large standard error due to small sample sizes. But another interpretation is that measures like population size (or industrial diversity) alone do not take into account the specific combinations of people and technology in a city that can increase (or decrease) the potential for capability recombination. {This is consistent with the findings by \citep{balland2017geography}, for example, who found that high-complexity patenting tends to be geographically concentrated, yet not exclusively in the most populous cities; or by \citep{sarkar2020evidence}, who found that to understand urban economic performance the specific intra-city composition of economic and social activity and physical infrastructure should be taken into account, in addition to just population size. This suggests that our estimate of city collective knowhow may capture such intra-city composition. 
}

{Building on the previous point, our model differs from previous theories in the urban scaling literature regarding the prevalence of phenomena in cities (e.g., levels of employment in industries, number of homicides in a city in a year, cases of infectious diseases) \cite{Arbesman2009,PanEtAl2013DensityDriven,Bettencourt2013,Yakubo2014PhysRevE}, which are anchored only in the suitability of the city to foster the occurrence of a given phenomenon. In previous models, the suitability of a city is determined by the networks of interactions within the city. Thus, these previous approaches neglected the idea that different activities (more precisely, the individuals engaged in them) would respond differently to the same urban context. The broad argument from these previous theories stated that since interactions in a network scale faster than linearly with the number of agents, urban indicators that are the result of social interactions would also scale superlinearly with population size. As a consequence, the canonical urban scaling model \cite{Bettencourt2013} predicts that all measures of output per capita will scale with the square root of city population density, regardless of any consideration about spatial equilibrium or budget constraints (see electronic supplementary material~G). Our model, on the other hand, adds a mechanism in which only the right combination of factors leads to the creation of employment, and since industries differ in the factors required, they will scale differently.
Instead of emphasizing that interactions scale faster than the number of individuals in a network, our model emphasizes the diversity of types of interactions growing with the number of interactions. The collective knowhow, $r$, quantifies what really matters about the context in which individuals are embedded: that some contexts are more complementary to some individuals and to some activities than others.}

{Our results have two main limitations, one theoretical and one empirical. First, our model does not specify how places acquire capabilities, or how capabilities emerge or evolve. {Hence, we currently lack a clear specification for how our model predicts economic change. That said, preliminary analysis suggests that our complexity variables are promising in explaining economic growth in cities (see electronic supplementary material~M for an analysis of how our estimate of collective knowhow correlate with GDP per capita level and  growth).} Incorporating time to our model} is part of future work, and one crucial insight can be drawn from the regression analyses we have presented here linking our model with the larger literature on economic complexity: we found that the \emph{number} of capabilities (present in cities or required by industries) are as important as the \emph{type} of capabilities. This result was found by comparing the regression coefficients of ECI and PCI proposed in \cite{HidalgoHausmann2009} against our ``drivers'' (column 7 in Tables~\ref{tab_firmsize_regs} and \ref{tab_wage_regs}, respectively). As is observed in those regressions, the inclusion of the ECI and PCI is statistically orthogonal to our measures of collective knowhow and industry complexity. If assume our model is correct, this finding is inconsistent with the conventional interpretation of the ECI and PCI. However, the fact that they are still statistically significant provides evidence that these quantities are capturing different information about economic activities in places. Namely, our estimates measure information about the number of capabilities in places and industries, while the PCI and ECI capture the specialization patterns about which types of capabilities are present in a place, or required by an industry (consistent with what is argued by \citep{mealy2019interpreting}). Hence, the important conclusion here is that both the number and the type of capabilities affect economic performance. The generalization of the model must thus require an explicit inclusion of technological similarities between industries as has been studied in the literature of related variety and related diversification \cite{boschma2011emerging,boschma2017relatedness,Neffke2013SkillRelatedness,boschma2011technological,frenken2007related,frenken2007theoretical}, and presumably include migration dynamics between cities to model the flow of capabilities (e.g., \cite{bahar2014neighbors,neffke2013agents,boschma2017neighbour,jara2018role}). {As for the empirical limitation, we must highlight that we have tested the validity of our model using data from US metropolitan areas. While the US urban system has been predominantly the focus of study for understanding cities, there is evidence that the regularities found in it are not necessarily universal (see, e.g., \cite{chauvin2017different,venables2017breaking}). Hence, applying this model and the method of estimating economic complexity metrics to explain urban outcomes in other parts of the world stands as an important avenue for future research.}

We believe our work contributes to the literature on economic complexity by providing conceptual and theoretical improvements that lead to a deeper understanding of cities as heterogeneous places where diverse individuals engage in an interconnected network of complex phenomena.


\section*{Competing interests}
We declare we have no competing interests.

\section*{Funding}
No funding has been received for this article.

\section*{Acknowledgment}
We thank F. Neffke, R. Hausmann, D. Diodato, C. Bottai, U. Schetter, and participants at the Growth Lab seminar at the Center for International Development, for helpful discussions and comments.


\bibliographystyle{RS}
\bibliography{Refs}
\newpage

\begin{appendices}
\input{Appendices.tex}
\end{appendices}

\end{document}

%% file: Appendices.tex
\setcounter{page}{1}
\setcounter{figure}{6}  
\setcounter{table}{2}  

\noindent{\Large{\bfseries Supplementary Material}} 
\vspace{0.5cm}

\noindent{\Large{\bfseries Estimating the drivers of urban economic complexity and their connection to economic performance}}
\vspace{0.5cm}

\noindent{\Large{ Andres Gomez-Lievano${}^{1,2}$ and Oscar Patterson-Lomba${}^{2}$}}\\
\small{${}^{1}$ Growth Lab, Harvard University, Cambridge MA, USA}\\
\small{${}^{2}$ Analysis Group, Inc., Boston MA, USA}

\vspace{1cm}

This document makes references to sections, equations, figures, tables, and citations in the main text. The numbering of figures and tables here continue that of the main text. 

\section{Derivation of the model}\label{app:modelderivation}

For mathematical and statistical convenience, assume the factors are ``assembled'' within or around individuals, but keep in mind that the ``assemblers'' of the elements can be households, firms, or other organizations. For example, a person can become an inventor not only if her encounter with another person is conducive to patenting, but if a large confluence of factors during her life happen in the right order in the right moment at the right place. To model these situations from the point of view of this theoretical framework we represent a given phenomenon by a list of ``factors'' or ``ingredients'', an individual by a vector of size equal to the list of ingredients flagging (e.g., identifying with 0s or 1s) which of the factors or ingredients she is already endowed with, and a city is represented by another vector (of the same size) flagging which of the factors or ingredients can be found in the city. A phenomenon occurs when the requirements for the phenomenon are all satisfied, either because the individual has them or because they are provisioned by the city.

Let us assume that a specific industry $f$ requires the combination of $M_f$ different and complementary capabilities. These may include knowhow of finance and accounting, a legal team, engineering capabilities, a team of technicians doing research and development, plus sales and marketing capabilities. Thus, we will typically think of ``capabilities'' as ``professional or job occupations'', although they can also include public services that a production process may need as a necessary requirement.

The parameter $M_f$ represents, in this view, the ``inherent complexity'' of the economic activity associated with the production of industry's product $f$. The more capabilities are needed, the larger the value of $M_f$, and the more complex the activity. Notice we are assuming that capabilities do not substitute each other, and the number of capabilities is large, i.e., $M_f \gg 1$. Thinking probabilistically in this model will provide several insights, and this is enabled by the assumption about the large multiplicity of capabilities \cite{gomez2016explaining}. 

Now, let $s_i$ represent a measure of how many of the capabilities individual $i$ already has. Specifically, let it be the probability that she has \emph{any} capability of the $M_f$ capabilities required by the typical business in industry $f$. This probability can be interpreted as a measure of her individual knowhow. For example, if the industry in consideration is shoe manufacturing, $s_i$ represents the chances she knows any one of the capabilities in a shoe manufacturing firm. The larger the parameter $s_i$ is, the better equipped she will be in engaging in the shoe manufacturing business, and, as we will see, the less there will be a need to put together a team of people to run such a business. Notice, however, that while $s_i$ can be interpreted as the level of schooling or education, it does not capture the \emph{depth} of knowledge but the \emph{breadth}: It is about how many different things she could know how to do individually. The probability she will get a job in a specific industry $f$ on her own merits is the probability she will have all capabilities, which is given by $s_i^{M_f}$. Since $s_i$ is a number between 0 and 1, the more complex the economic activity, the probability she will be successful finding a job will decrease exponentially with $M_f$. This probability, however, does not yet account for the fact that she lives in a city and the probability of being employed actually depends on finding a space (e.g., a place of work) that already has people with the complementary capabilities she does not have. 

For this, suppose the city $c$ ``provides'' $D_c$ capabilities to individual $i$ (where $0\leq D_c\leq M_f$). In other words, through her exposure to other sources of capabilities from living in city $c$, like family, friends, colleagues or, in general, public and private services, individual $i$ could in principle be able to get and complete missing skills and capabilities outside her expertise. Presumably, the bigger the city, the more diverse, and the larger $D_c$ will be. 

Since $D_c$ are the capabilities provided by the city, a job in business $f$ in city $c$ requires that individual $i$ knows $M_f - D_c$ capabilities. These are capabilities that she will need to bring to the business on her own, without the help of the city. Thus, the probability that $i$ will be employed in industry $f$ given that she lives in a city where she has access to $D_c$ capabilities is equal to
\begin{align}
	\Pr(X_{i,c,f}=1~|~D_c)&= s_i^{M_f-D_c}.
	\label{eq_probgivenDc}
\end{align}
Thus, living in a diverse city has the effect that finding a job is exponentially easier.

In reality, however, $D_c$ is not a fixed number. Cities are dynamic places, they change from neighborhood to neighborhood and from day to day, and no person is exposed to the city as a whole. Hence, if individual $i$ is very unlucky she may get $D_c=0$, or she can be super lucky and get $D_c=M_f$. To take this stochasticity into account, we can also think instead of the \emph{probability} that the city provides \emph{any} of the capabilities. Let us denote this probability by $r_c$. The expected number of capabilities offered in the city, required to get a job in $f$, is $\mathrm{E}[D|\text{city $c$}] = r_c M_f$. Thinking of $D_c$ probabilistically, means thinking of $D_c$ in this context as a ``binomially distributed random variable'' $D$ with parameters $M_f$ and $r_c$. 

To correctly compute the probability that individual $i$ will get a job, we thus need to average Eq.~\eqref{eq_probgivenDc} over all the possible number of capabilities the city may offer:
\begin{align}
	\Pr(X_{i,c,f}=1) &= \sum_{D_c=0}^{M_f}\Pr(X_{i,c,f}=1~|~D=D_c)\Pr(D=D_c) \nonumber \\
	&= \sum_{D_c=0}^{M_f}s_i^{M_f-D_c}\binom{M_f}{D_c}r_c^{D_c}(1-r_c)^{M_f-D_c} \nonumber \\
	&= \sum_{D_c=0}^{M_f}\binom{M_f}{D_c}r_c^{D_c}(s_i(1-r_c))^{M_f-D_c} \nonumber \\
	&= (r_c + s_i(1-r_c))^{M_f} \nonumber \\
	&= (1 - (1-s_i)(1-r_c))^{M_f}.
	\label{eq_prob}
\end{align}

We can generalize this model and imagine that the city has a ``field'' spread in the $xy$-coordinates, $r_c(x,y)$. This field is an abstraction of the urban milieu, it represents the probability that the city provides one of the ingredients for phenomena to occur, and we can assume that people interact with it as they live and work in the city. It should capture the elements from all the types of urban interactions to which people are exposed (both the social and built environment). In this view, the city functions as though it is permeated across space by a ``cultural field'', and $r_c(x,y)$ quantifies the magnitude of the social, economic, and cultural repertoire available at a particular location. When the value of the field is high, it means that this location in the city has a high concentration of ``diverse urban factors''. A high value of $r_c(x,y)$ will therefore increase the probability that an individual will find the right elements to engage in a given urban phenomenon (e.g., to be employable in an industry). Eq.~\eqref{eq_prob} assumes individuals interact with the average intensity of the field, $r_c\equiv\langle r_c(x,y)\rangle$, where $\langle \cdot\rangle$ is a spatial average. The mean field approach allows us to isolate a core mechanism in the model: that urban phenomena occur because individuals are able to ``execute'' a recipe (e.g., a production process, a program or algorithm) if the environment is favorable, that is, if the city complements the individual. How complex a given recipe is, how capable is an individual, and how suitable is the city for executing the recipe are the three fundamental quantities that determine the overall statistics of employment in cities, as well as other measures of urban output.

Equation~\eqref{eq_prob} implies an exponential function. With some minor rearrangement, the probability that $X_{i,c,f}=1$ can be written as
\begin{equation}
	\Pr\{X_{i,c,f}=1\}=e^{-M_f (1-s_i)(1-r_c)}. \label{epercapitaapp}
\end{equation}
The exponent is the product of three quantities, respectively associated with the phenomenon, the person, and the city. These are the drivers of employment in the city, which is why we refer to them as the drivers of urban economic complexity. We will flesh out the meaning of each of these terms below, but we want to emphasize that the value of such equation is that, through it, the model establishes null expectations regarding the broad patterns manifested across urban phenomena. Such approach is typical in ``mean field theories'' in physics.

\section{Complementarity between urban and individual knowhow}\label{app:complementarity}
Our model does not account for the forces of markets and prices. However, we offer here some simple arguments to show that our model does predict assortative matching between skilled individuals and complex cities, for certain combinations of the parameters. 
		
		To see this, recall our function $p(M, s, r)=\exp\left(-M(1-s)(1-r)\right)$ (this is the main expression of our model, the probability that an agent with individual knowhow $s$ is employed in a sector of complexity $M$ in a city with a level of collective knowhow $r$). Suppose $M$ is not a choice and is fixed. Assume there are two individuals, one with many skills $s_H$ and the other with few skills $s_L$. Assume, in addition, two cities, one with a high level of collective knowhow $r_H$ and the other with a low level $r_L$. The question is: if there are congestion costs (i.e., both individuals do not want to go to the same city), how will individuals sort themselves? What will maximize the total probability of employment? Will the high-knowhow individual prefer to go the low-knowhow city? Or will there be assortative matching ($s_H$ with $r_H$, and $s_L$ with $r_L$)? 
		
		Assortative matching will happen if $$p(M, s_H, r_H) + p(M, s_L, r_L) > p(M, s_L, r_H) + p(M, s_H, r_L).$$ This condition holds if the cross-derivative $\frac{\partial^2 p(M, s, r)}{\partial s \partial r}>0$. In other words, when the function $p(M, s, r)$ is \emph{supermodular}. According to the cross-derivative of our function $p(M, s, r)$ (not shown), we conclude it is supermodular when the condition $$(1-s)(1-r)>1/M$$ is met. Thus, in the limit of highly complex industrial activities $M\gg 0$, the right hand side will tend to zero $1/M\rightarrow 0$, and the condition will be met. Therefore, the individual knowhow of workers will be matched assortatively with the collective knowhow of cities (since $r$ and $s$ are positive and strictly less than unity). 
		
		\begin{figure}[h]
			\centering
				\includegraphics[width=0.5\textwidth]{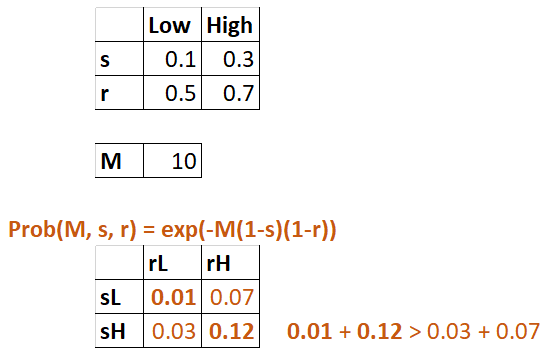}
			\caption{Example of assortative matching between individuals and cities.}
			\label{fig:example1}
		\end{figure}		
		Figure~\ref{fig:example1} shows an example in which it is clear that  $p(M, s_H, r_H) + p(M, s_L, r_L) > p(M, s_L, r_H) + p(M, s_H, r_L)$ for $M=10$, $s_L=0.1$, $s_H=0.3$, $r_L=0.5$, and $r_H=0.7$.

\subsection{When does employment scale superlinearly with population size?}
Taking as a given the functional form of $p(M, s, r)$, we can evaluate the conditions in which $\partial \ln(p)/\partial \ln(N) >0 $, assuming $r$ and $s$ (but not $M$) are functions of $N$.

Taking derivatives, we arrive at the following condition:
\begin{align}
	\frac{\partial}{\partial \log(N)}\log(1-s) + \frac{\partial}{\partial \log(N)}\log(1-r) < 0.
\end{align}
Assume, for simplicity, that $s$ and $r$ are related to population size $N$ as $1-s = a_0 N^{-a_1}$ and $1-r=b_0 N^{-b_1}$, with coefficients $a_i,b_i$ fixed. This implies that shares of employment will increase superlinearly if $$a_1 + b_1 > 0.$$

\section{Relative impact of the drivers on economic performance}\label{app:impactsofdrivers}


Small changes in any of the three terms can in principle have very large effects on the employability of a person. If we denote $p_{i,c,f}$ the probability in Eq.~\eqref{epercapitaapp}, the change in each term has the following meaning:
\begin{itemize}
	\item Technological improvement of production process of $f$:
	\begin{align}
		\frac{\partial p_{i,c,f}/\partial (-M_f)}{p_{i,c,f}}&= (1-s_i)(1-r_c), \label{eq_Mp}
	\end{align}
	
	\item Individual learning for individual $i$:
	\begin{align}
		\frac{\partial p_{i,c,f}/\partial (M_f s_i)}{p_{i,c,f}}&= ~ (1-r_c), \label{eq_si}
	\end{align}
	
	\item Collective learning for city $c$:
	\begin{align}
		\frac{\partial p_{i,c,f}/\partial (M_f r_c)}{p_{i,c,f}}&= ~ (1-s_i) \label{eq_rc}.
	\end{align}
\end{itemize}
The partial derivatives have the term $M_f$ because we want them to reflect changes in the \emph{number of capabilities}, not changes in the parameters themselves. Using the same units of change allows us to compare these rates. Hence, $\partial (-M_f)\equiv-\partial M_f$ represents the reduction of the number of capabilities required by industry $f$, $\partial (M_f s_i)\equiv M_f\partial s_i$ represents the increase in the average number of capabilities known by the individual $i$, and $\partial (M_f r_c)\equiv M_f \partial r_c$ represents the increase in the average number of capabilities present in city $c$.\footnote{We interpret a reduction in the number of capabilities $M_f$ required by industry $f$ as a technological improvement because we associate it to a sophistication in physical capital. This sophistication occurs when tasks are bundled, automated and simplified, and thus we expect this process to imply a reduction in the number of capabilities required by a production process.} In this way, the probability that individual $i$ will be employable, $\Pr\{X_{i,c,f}=1\}$, will increase according to Eq.~\eqref{eq_Mp} through technology improvements, Eq.~\eqref{eq_si} tells us that it will increase through individual learning, and Eq.~\eqref{eq_rc} that it will increase through collective learning.

\begin{figure}[!t]
	\begin{center}
		\includegraphics[width=0.98\textwidth]{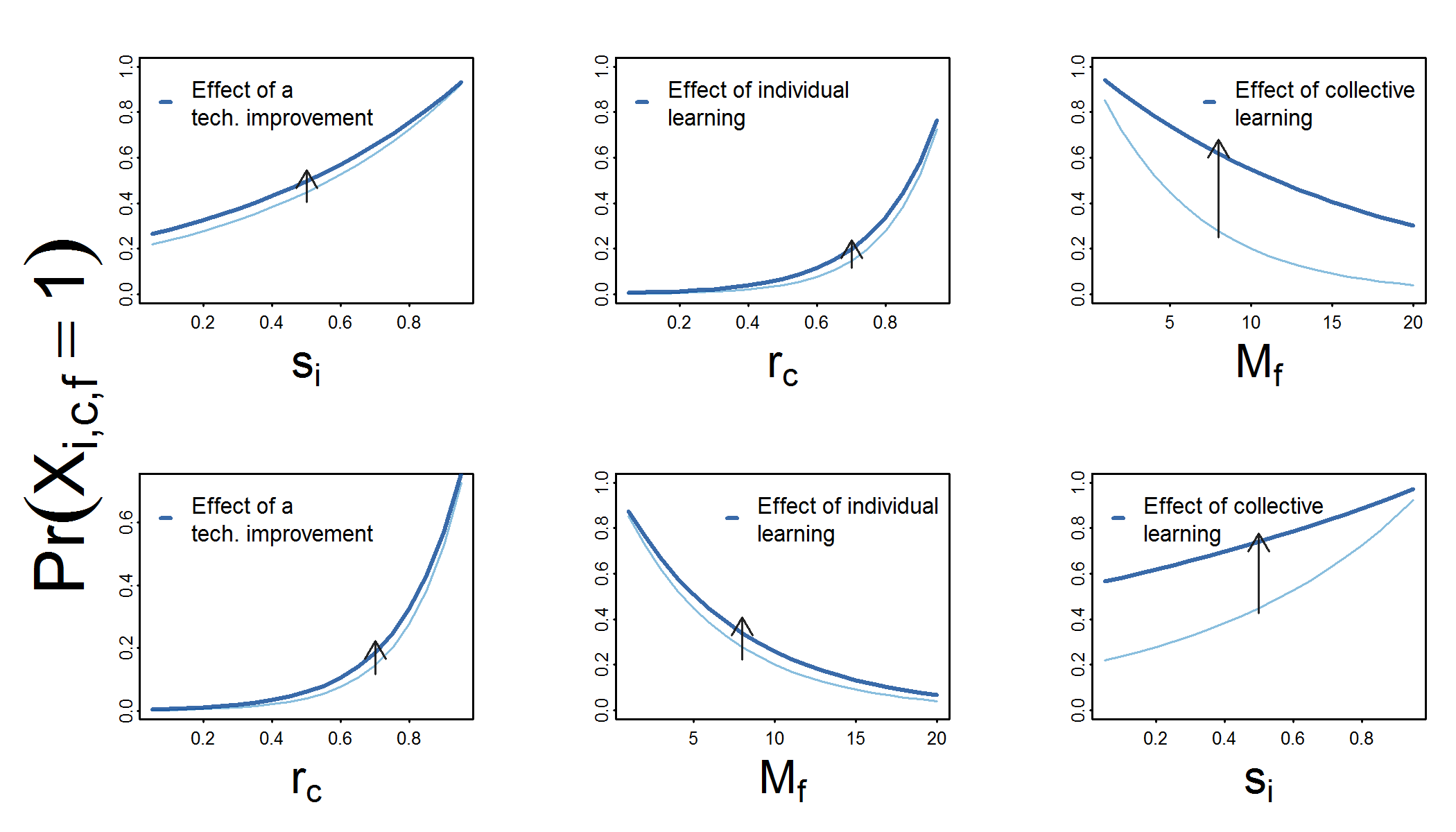}
		\caption{Comparing the three ways of increasing the probability that individual $i$ in city $c$ is employed in industry $f$. For each panel, one of the parameters is explicitly shown to vary across the $x$-axis, another parameter is changed in order to represent the change in probability (either by a technological improvement in the left panels, individual learning in the middle panels, and collective learning in the right columns), and another parameter is implicitly kept constant, correspondingly at values $M_f=8$, $s_i=0.2$, or $r_c=0.8$. In each panel, the change from lightblue to darkblue lines represents the increase in probability due to a change in $M_f$ (left panels), $s_i$ (middle panels), and $r_c$, equivalent to one capability. Hence, a technological improvement is when $M_f$ is reduced by 1, individual learning is when $s_i$ is increased by $1/M_f$, and collective learning is when $r_c$ is increased by $1/M_f$. }
	\label{fig_model_learning}
	\end{center}
\end{figure}

Since the city is seen here as a large set of capabilities, the probability it provides any input is much larger than the probability an individual has it, so $r_c\gg s_i$. Conversely, $1-r_c\ll 1- s_i$. Consequently, we have that $0 < (1-r_c)(1-s_i) < 1-r_c \ll 1 - s_i$. The implication is that these rates have the following order:

\begin{align}
	0<\frac{\partial p_{i,c,f}/\partial (-M_p)}{p_{i,c,f}} < \frac{\partial p_{i,c,f}/\partial (M_p s_i)}{p_{i,c,f}} < \frac{\partial p_{i,c,f}/\partial (M_p r_c)}{p_{i,c,f}}
	\label{eq_comparison}
\end{align}
Thus, the effect on the probability of employment of a technology improvement is smaller than the effect of individual learning which is smaller than the effect of collective learning. Increasing the collective knowhow of city $c$ can occur through immigration that brings capabilities and knowhow in people not available already in $c$\footnote{Models such as that in \cite{borjas2014immigration} show that when skills of immigrants are complementary to the locals wages of both locals and immigrants increase.}, direct foreign investments that inject specific capabilities to specific industries, or by pure innovation. These have a significant effect on the probability that $i$ will be employed in $f$ according to Eq.~\eqref{eq_comparison}. Figure~\ref{fig_model_learning} illustrates these effects separately.

Presently, we lack a detailed theory about the dynamical laws of these drivers and how they relate to one another. Still, the comparisons in Eq.~\eqref{eq_comparison} allow us to remark that increases in collective knowhow may have a reinforcing effect which initiates a virtuous cycle: a place with a relative large body of collective knowhow will attract more people and facilitate more complex economic activities, which themselves will increase the body of collective knowhow in that place. This process will thus propel a run-away cycle of collective learning that will concentrate economic activities and wealth in large cities (see, e.g., \cite{OClery2018}). The more complex the activities, the more concentrated they will be in very few places. Once again, this explains why complex activities (e.g., being an inventor) tend to happen and concentrate disproportionately more in large cities as compared to less complex activities \cite{gomez2016explaining}.

The comparison in Eq.~\eqref{eq_comparison} hinges on highly simplifying assumptions. For example, the comparison assumes that a linear (infinitesimal) change in the three variables is comparable among them. In other words, it does not take into account the \emph{cost} of these changes. Furthermore, the model does not take into account externalities, like the effect that a technological change in industry $f$ has on an industry $f'$. Nevertheless, with these comparisons one can hypothesize which of the three drivers has the largest impact on economic performance.

\section{Data sources}\label{app:datasources}
Data was downloaded using the Bureau of Labor Statistics API through the website \url{http://www.bls.gov/cew/doc/access/data_access_examples.htm}. We use data on the estimated counts of employment, number of establishments and average wages by city-industry-year. The reader should be aware that `establishments' and `firms' do not refer to the same thing. Establishments are typically synonymous to `plants', and firms can be multi-establishment: ``An establishment is a single physical location where one predominant activity occurs. A firm is an establishment or a combination of establishments'' (from \url{https://www.bls.gov/opub/mlr/2016/article/establishment-firm-or-enterprise.htm}). 

The specific data for cities are the definition of metropolitan statistical areas which was selected using the guide in \url{http://www.bls.gov/cew/doc/titles/area/area_titles.htm}. We will use the terms metropolitan areas and cities interchangeably. The metropolitan codes are from the 2004 definitions, but we drop the metropolitan areas located in Puerto Rico. The list of industry codes can be found in \url{http://www.bls.gov/cew/doc/titles/industry/industry_titles.htm}. We use employment numbers aggregated to 3-digit level industries.

The original industry 3-digit classification has 91 different industries. However, we cannot blindly apply our theory to all industries. Our theory applies to phenomena that can be regarded as the \emph{output} of urban processes. This constrains the phenomena and the type of activities that we can analyze. In the case of industries, our model does not apply to those for which firms are constrained by the local demand for their product or service (e.g., ``Scenic and sightseeing transportation''), or to those that are natural-resource-based (e.g., ``Oil and gas extraction''). In other words, our model applies to the firms whose existence depends instead on whether there is availability of the required capabilities for their processes, and are firms that can in principle sell their products outside the city (i.e., demand can be assumed to be infinite). Hence, we kept in our analysis 78 industries that were likely to be less constrained by local demand for their products or services and more likely to be constrained by their production requirements (e.g., availability of high-skilled workers), and we discarded the other 23. See Table~\ref{tab_inds_excluded} for the list of excluded industries, and Table~\ref{tab_inds_included} for those we included.

To construct the measures of occupational diversity for industries, we use the yearly tables, from 2003 to 2016, that are published in May (e.g., \url{https://www.bls.gov/oes/special.requests/oesm16in4.zip}).

Data on Gross Metropolitan Product and Population can be downloaded from the Bureau of Economic Analysis.

\section{Estimating individual knowhow}\label{app:individualknowhow}

Assuming $s_i$ across individuals in city $c$ are statistically distributed according to probability $p_S(S=s_i|c)$, we want to compute the expected mean of $\Pr\{X_{i,c,f}=1\}$ across individuals in $c$. Assuming the values of $s_i$ do not differ significantly from an average $\bar{s}_c$ (i.e., $S$ approximates a degenerate random variable), we can use Jensen's inequality such that the arithmetic mean can be approximated by the geometric mean. Hence,
\begin{align}
	\Pr\{X_{c,f}=1\} &= \mathrm{E}_S\left[\Pr\{X_{i,c,f}=1\}\right] \nonumber \\
	&= \int_0^1 e^{-M_f (1-s) (1-r_c)} p_S(s|c) \mathrm{d}s \nonumber \\
	&\approx e^{-M_f (1-r_c)\int_0^1 (1-s) p_S(s|c) \mathrm{d}s}\nonumber \\
	&= e^{-M_f (1-r_c)\mathrm{E}_S[1-s|c]}\nonumber \\
	& \approx e^{-M_f (1-\bar{s}_c) (1-r_c)},
\end{align}
in which we obtain the same functional form as before, but instead of $s_i$ we have the average individual knowhow in city $c$, $\bar{s}_c \approx \sum_{i\in c}s_i/N_c$.

In the construction of the model, we emphasized that the quantity $s_i$ captures the competency of individual $i$ to be endowed with any given capability. In the context of the model, a large value of $s_i$ means that individual $i$ will be able to know how to do many different things. We do not have a direct way of estimating this quantity.

Given the regression model we propose, we only have $\widehat{\gamma_c}$, which is an estimate of $-\ln((1-\bar{s}_c)(1-r_c))$. In order to separate $\widehat{\gamma_c}$ into the individual-level and city-level quantities, we will assume that $s_i$ is positively correlated with educational attainment. More years of educational attainment are usually a measure of specialization. But they are also an indication of a person's competency, and thus we will proxy $\bar{s}_c$ using the average levels of schooling in the city $c$. 

The American Community Survey (ACS) provides data on Educational Attainment through the website \url{https://factfinder.census.gov/faces/tableservices/jsf/pages/productview.xhtml?pid=ACS_16_5YR_S1501&prodType=table}. We use the data corresponding to the 5-year averages from 2009 to 2016. For the ``Population 25 years and over'', we collect the population numbers for the seven different levels of educational attainment available: Less than 9th grade, 9th to 12th grade (no diploma), high school graduate, some college (no degree), associate's degree, bachelor's degree, and graduate or professional degree. We use an ordinal variable for each level, from 1 to 7, and we average them weighting by the number of people counted with that educational level:
\begin{align}
	\widehat{\bar{s}_c} = \sum_{l=1}^7 l\times \frac{P_c(l)}{P_c},
\end{align}
where $P_c$ is the size of the population 25 years or older in city $c$, and $P_c(l)$ is the portion whose maximum level of education is level $l$. By construction, $\sum_l P_c(l) = P_c$.

\section{Alternative models}\label{app:alternativemodels}
As it is shown in the text, Equation~(\ref{epercapitaapp}) suggests a methodology to estimate the parameters $M_f$, $r_c$ and $s_i$. If one lacks individual-level microdata, and has data only aggregated at the level of employment by industry and city, one can still estimate $M_f$ and a city-specific variable that stands for $(1-\bar{s_c})(1-r_c)$, where $\bar{s_c}$ is the average $s_i$ across individuals in city $c$. 

The way to estimate is by taking logarithms twice to the shares of employment across industries in a city, making sure to multiply by negative 1, $-\log(-\log(p_{c,f}))$. After this transformation, one can run a regression with industry and city fixed-effects. 

What happens, however, if we change the assumption regarding capabilities, from ``complementary'' to ``substitutable''? This change leads to the prediction that one should instead estimate city knowhow and industry complexity with fixed effects in a regression on $\log(p_{c,f})$ (instead of $-\log(-\log(p_{c,f}))$):

To see this, assume, instead, that capabilities are substitutable. That is, person $i$ is employed if she has \emph{any} of the $M_f$ capabilities required by the industry (as opposed to having to get \emph{all} capabilities). As before, capabilities can come from the city or from herself. The way to compute $\Pr(X_{i,c,f}=1)$ under the assumption of substitutable capabilities is instead to calculate $\Pr(X_{i,c,f}=0)$. 
		
		The probability of not getting employed ($X_{i,c,f}=0$) is the probability that she gets none of the capabilities. The probability of not getting one particular capability is $(1-s_i)(1-r_c)$. The probability of not getting any of the $M_f$ capabilities is thus $[(1-s_i)(1-r_c)]^{M_f}$. Thus, the probability of getting employed would be one minus that: $\Pr(X_{i,c,f}=1)=1 - [(1-s_i)(1-r_c)]^{M_f}\approx M_f(s_i + r_c)$. Finally, assuming workers in city $c$ have an average individual knowhow $\bar{s_c}$, we get that a random worker in city $c$ will be employed in industry $f$ with probability $p_{c,f}=M_f \tau_c$, where $\tau_c=\bar{s_c} + r_c$ is a city-specific effect. From this, we conclude that if the world follows a logic in which capabilities are substitutable, the model we called ``Model 1.2'' in which we regress $\ln(y_{c,f})=\delta_f + \gamma_c$ should outperform our model.

Note that the model in Equation~(4.2) in the main text has a free parameter for each industry and one for each city, hence it is more complex (in terms of degrees of freedom) than alternative models that may consist of fewer parameters. Since our model has so many degrees of freedom, it can accommodate and fit well data. Therefore, it is not wise to falsify the theory based on the data we used to fit the model. Our model, in other words, has the risk of overfitting the data. The hypothesis to test is whether our model has a superior capacity to predict values of $y_{c,f}$ which were not used in the model-fitting procedure (``out-of-sample prediction''), as compared to other alternative models. If our model is indeed comparatively better, then we will have evidence that the theory is capturing meaningful aspects of reality that other models are not \cite{shmueli2010explain}.

Hence, we propose to evaluate the predictive power of model Eq.~(4.2) of the main text with respect to the following alternative models using holdout data. These alternative models require less, or the same, number of parameters: 
\begin{description}
	\item[Model 1.1:] $$y_{c, f, t} = \delta_{f, t} + \gamma_{c, t} + \varepsilon_{c,f,t}.$$
	\item[Model 1.2:] $$\ln\left(y_{c, f, t}\right) = \delta_{f, t} + \gamma_{c, t} + \varepsilon_{c,f,t}.$$
	\item[Model 2.1:] $$\ln\left(y_{c, f, t}\right) = \alpha_{f, t} + 0.16 \ln\left(N_{c, t}\right) + \varepsilon_{c,f,t}.$$
	\item[Model 2.2:] $$\ln\left(y_{c, f, t}\right) = \alpha_{f, t} + \beta_{f,t}\ln\left(N_{c, t}\right) + \varepsilon_{c,f,t}.$$
\end{description}
Note we have divided models in two types: fixed-effects models that differ among themselves in the functional transformation of the dependent variable\footnote{We thank F. Neffke for suggesting these alternatives.}, and urban scaling models that use population size as the independent variable. In other words, models $1.1$ and $1.2$ change the left-hand side of the regression equation while models $2.1$ and $2.2$ change the right-hand side. The subscript $t$ to denote the year will be dropped for simplification in what follows, unless needed explicitly.

Model $1.1$ assumes the per capita rates of employment are driven by two additive terms, one from the industry and the other for the city. Model $1.2$ assumes, instead, that the per capita rate is the product of the industry and city fixed-effects, and thus by taking logarithms we separate such interaction. Notice that in models $1.1$ and $1.2$, as in our model (4.2), we exclude the intercept. Differences in performance between model (4.2) and models $1.1$ and $1.2$ will inform us about the importance of the specific functional form predicted by the theory.

Model $2.1$ is the standard urban scaling model where it is assumed that the scaling exponent is the same for all phenomena, as suggested by network-based explanations. We will use population as the measure of city size. Results of our analysis, however, are qualitatively the same if one uses alternative measures like total employment. Model $2.2$ is an unconstrained version of Model $2.1$, where we assume that both the baseline prevalence and the scaling exponent differ, in principle, for each industry $f$. Differences in performance between model (4.2) and models $2.1$ and $2.2$ will inform us about the validity of adding degrees of freedom to explain employment patterns across cities and industries.

For each year, we will split the data into training and testing sets. These sets are defined as lists of randomly chosen pairs $(c, f)$ of cities and phenomena. Some of these pairs will either belong to ${\mathcal R}$ (the train set) while others to ${\mathcal S}$ (the test set). After the parameters of the models are fitted using the training data, they are then compared by how accurately they predict the dependent variable on the test set. We choose $20\%$ for the size of the test set.\footnote{Another approach to compare models would be to fit the models on all the data available and then use some goodness-of-fit statistic penalizing by the complexity of each model, like the Akaike Information Criterion (AIC) or the Bayesian Information Criterion (BIC), and then choosing that model which performs best (e.g., the model with the lowest AIC value). This approach is in general asymptotically equivalent to the presented above (see, e.g., \cite{stone1977asymptotic,fang2011asymptotic}). The ``train-then-test'' approach was chosen for two reasons. First, the quantity of data is large enough that it allows to have large sizes for both train and test sets for doing cross-validation. This frees us from the underlying assumptions behind metrics like AIC or BIC. Second, and more importantly, out-of-sample prediction is a fair and assumptions-free method to compare different models (see \cite{friedman2001elements} for further discussion).} 

The predictions will be evaluated using 
the \emph{root mean squared error}, $${RMSE} \equiv \sqrt{\frac{1}{\left|{\mathcal S}\right|}\sum_{(c,f)\in{\mathcal S}}\left(y_{c,f} - \widehat{y_{c,f}}\right)^2},$$ and 
the \emph{mean absolute error}, $${MAE} \equiv \frac{1}{\left|{\mathcal S}\right|}\sum_{(c,f)\in{\mathcal S}}\left|y_{c,f} - \widehat{y_{c,f}}\right|,$$ 
where $\widehat{y_{c,f}}$ is the predicted value. Both metrics quantify the predictive accuracy of models, but $RMSE$ is more sensitive to outliers, while $MAE$ quantifies the average prediction errors weighting all deviations equally. $MAE$ is thus preferable over $RMSE$, although we show both for completeness. The train and test random splits will be repeated $100$ different times of the data (bootstrapping cross-validation).\footnote{Notice this exercise is different from a typical predictive exercise using machine learning techniques, which may split the data in three (or more): a training set used to fit models, a validation set used to optimize models and select any free ``hyper-parameters'' they may depend on, and a test set that is used only once to report and compare final performance across models. Our models do not have hyper-parameters and, as a consequence, we only need two-way splits of the data.}

\section{Canonical scaling with population density}\label{app:bettencourtimplication}
In the main text, we said that canonical network-based and density-based models of urban scaling have the limitation that they predict a \emph{unique} way in which \emph{all} phenomena will scale with population density. To see this, assume that a measure of output per capita (e.g., GDP per-capita) is proportional to the number of interactions $y\propto I$ between between a person and the rest of people in the city (or a fraction), as in \cite{Bettencourt2013}. The fraction of interactions, in turn, is proportional to the density of social interactions, $I\propto N/A_{\text{social}}$, because $A_{\text{social}}$ is the area in which social interactions occur (e.g., the infrastructural network). Assuming individuals are uniformly spread in an physical area $A_{\text{spatial}}$, then the average distance between individuals is $d$, and we get that the area is proportional to $A_{\text{spatial}} \propto N d^2$. Assuming social interactions in space span the ``social area'' defined by the tree of close proximity interactions (see \cite{SamaniegoMoses2008,Banavar1999SizeAndForm}), one obtains that $A_{\text{social}} \propto N d$, which implies that $A_{\text{social}}\propto (N A_{\text{spatial}})^{1/2}$. If $\rho=N/A_{\text{spatial}}$ is the spatial population density, then we get that $y\propto I \propto N/A_{\text{social}} \propto (N/A_{\text{spatial}})^{1/2}$ which means $$y \propto \rho^{1/2}.$$ According to this reasoning, output per-capita scales with the square root of city population density. 

From this, Bettencourt (2013) \cite{Bettencourt2013} derives a direct association of output per capita with population size alone by proposing some constraints on the minimum budget and cost of traversing the city, which implies an association between the surface area of the city and population size. Hence, given the constraints, he gets that $A_{\text{spatial}}\propto N^{D/(D+H)}$, where $D$ is the dimensionality of the city (in principle, $D=2$) and $H$ is the fractal dimension of the path that people use to traverse the city. One could argue that different scaling exponents across phenomena are explained by the existence of different values of $H$ which apply to different economic activities or industries. However, we emphasize that differences in $H$ would still not change the association between output and population density, which is why we claim that network/density-based explanations predict a fix scaling with density, \emph{regardless} of any consideration about spatial equilibrium and budget constraints.

\newpage
\section{List of industries excluded and included in analysis}\label{app:listindustries}
\input{NHB_10001_industries_excluded.tex}
\newpage
\input{NHB_10001_industries_included.tex}

\newpage
\section{The effects of changing the dependent variable}\label{app:depvars}

We show in the main text that when comparing models that regress different forms of the dependent variable on city- and phenomenon-fixed effects, our model performs best. But what is the effect of alternating between the functions $y$, $\ln(y)$, and $-\ln(-\ln(y))$?

\begin{figure}[b!]
	\begin{center}
		\includegraphics[width=0.9\textwidth, trim = 0in 0in 0in 0in]{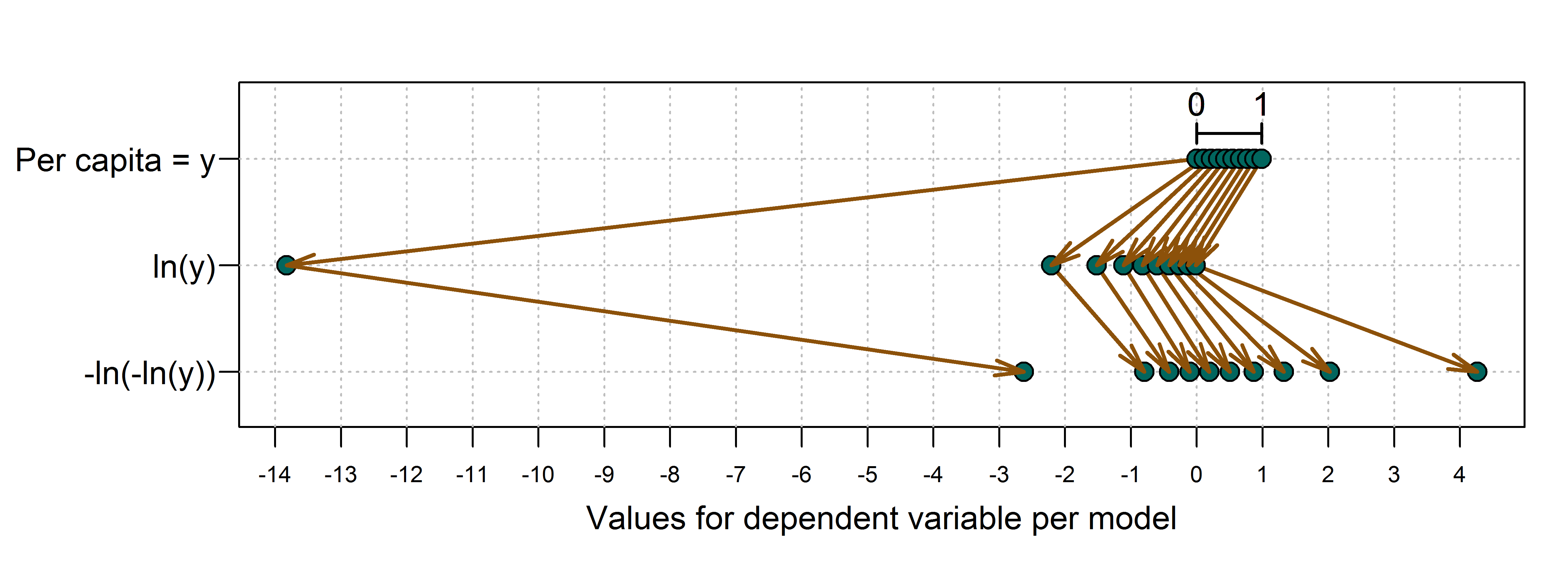}
		\caption{The effect of taking iterated logarithms on numbers between 0 and 1.} 
	\label{Fig_depvars}
	\end{center}
\end{figure}

\begin{figure}[b!]
	\begin{center}
		\includegraphics[width=0.9\textwidth, trim = 0in 0in 0in 0in]{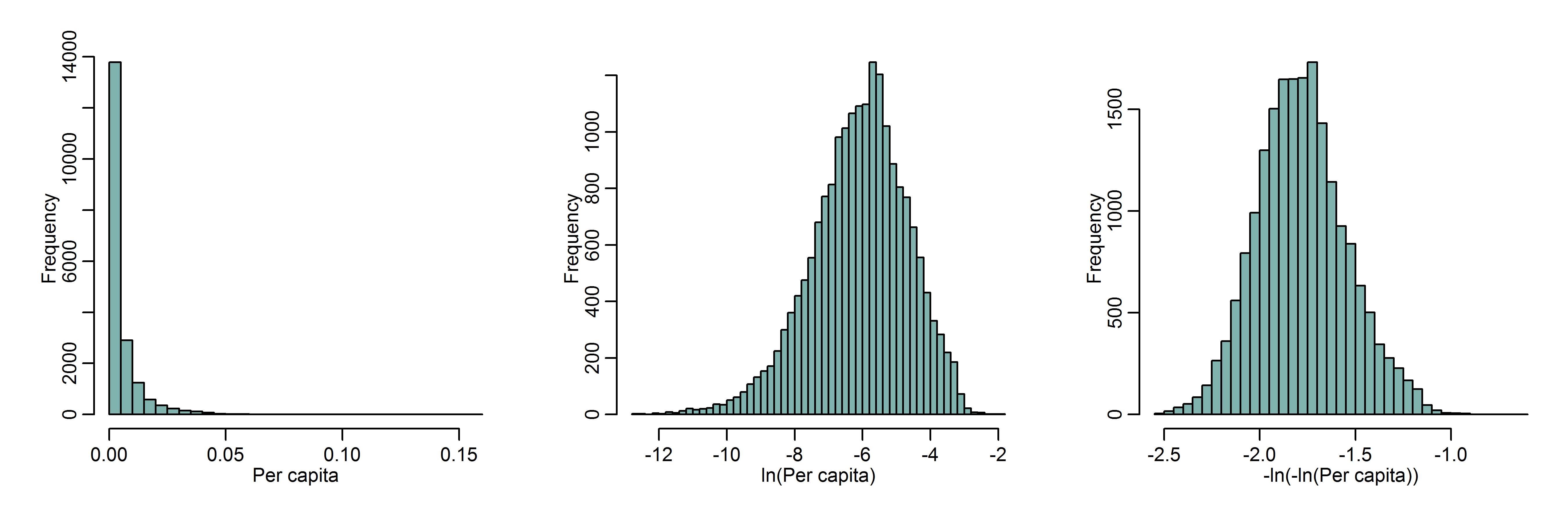}
		\caption{Histograms of values of the dependent variables $y$, $\ln(y)$, and $-\ln(-\ln(y))$ where $y$ is employment per capita across different industries and cities in the year 2016.} 
	\label{Fig_histdepvars}
	\end{center}
\end{figure}

Figure~\ref{Fig_depvars} shows that taking the logarithm of the per capita rates a single time amplifies the differences between the smallest values. Interestingly, taking the logarithm twice seems to amplify both small and large values, but to a less degree. Figure~\ref{Fig_histdepvars} shows the histograms of the per capita values in our data, according to the three transformations. This effect is what explains why the second best model is not to regress the logarithm of per capita rates on the fixed effects, but to regress directly the per capita values without taking logarithms.

\newpage
\section{Evolution of rankings of cities and industries according to drivers of urban economic complexity}\label{app:rankings}
\begin{landscape}
 \begin{figure}
  \centering
  \includegraphics[width=1.0\linewidth,keepaspectratio]{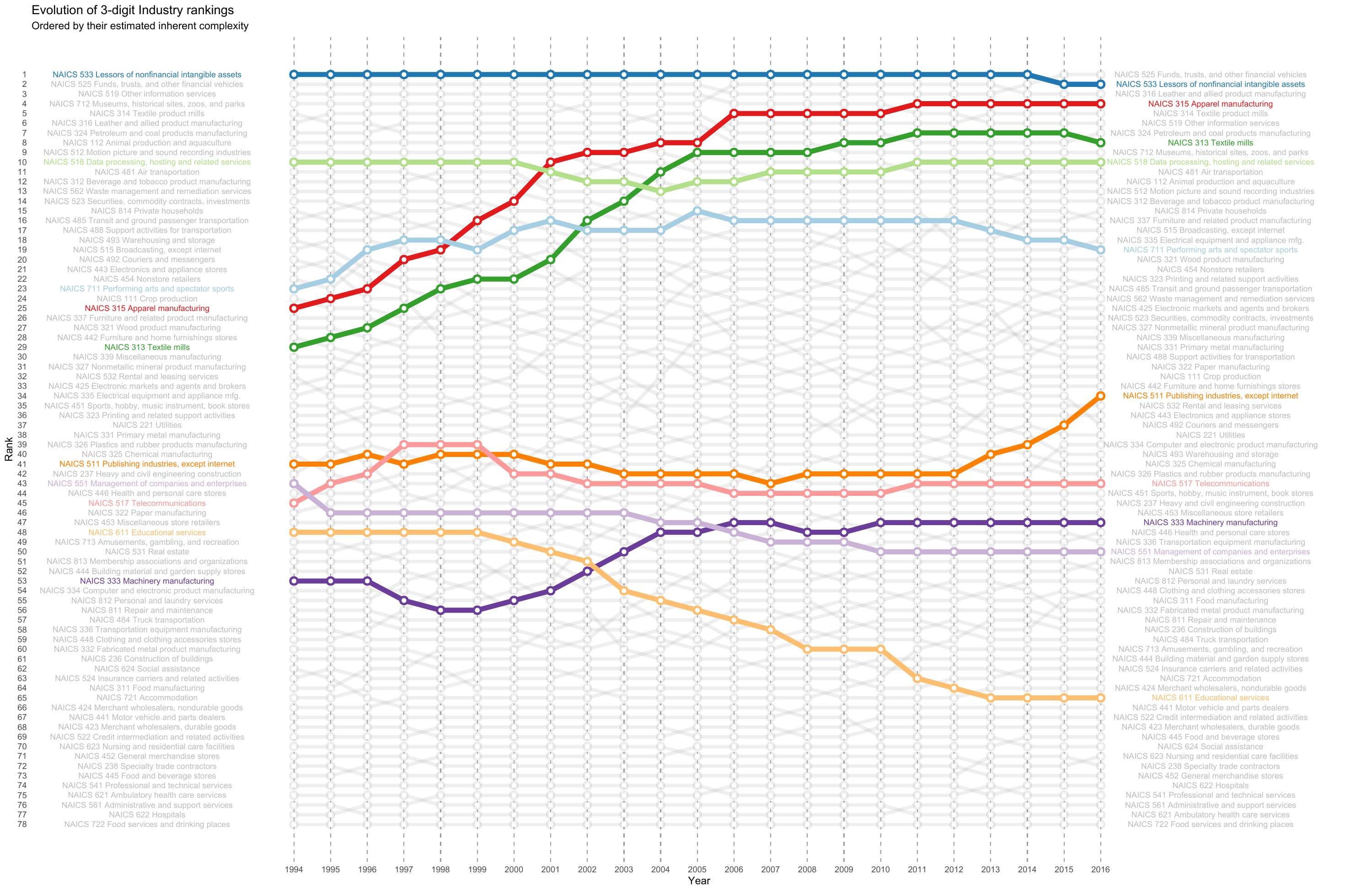}
  \caption{}
  \label{fig_ranking_complexity}
 \end{figure}
\end{landscape}

\begin{landscape}
 \begin{figure}
  \centering
  \includegraphics[width=1.0\linewidth,keepaspectratio]{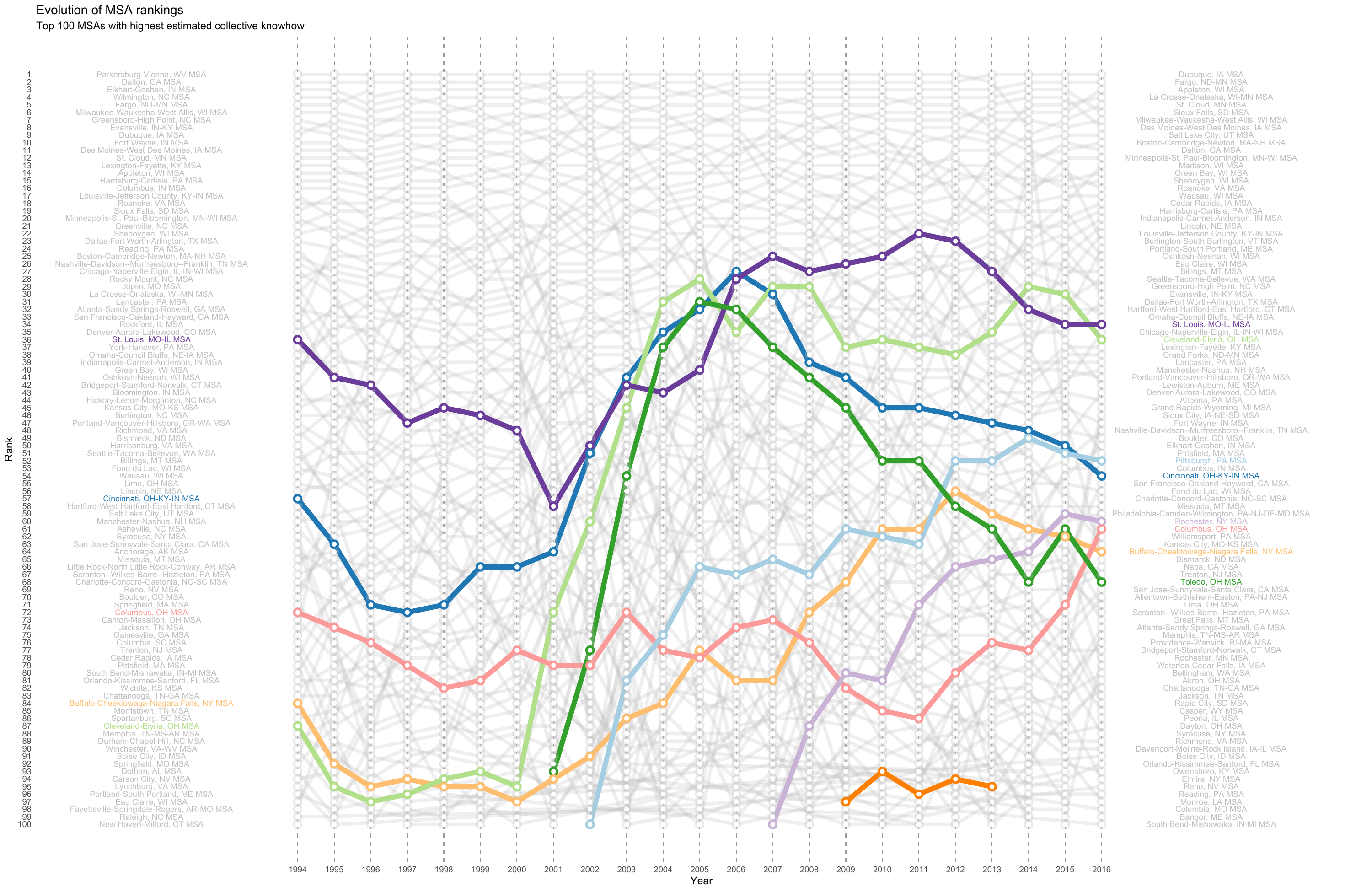}
  \caption{}
  \label{fig_ranking_knowhow}
 \end{figure}
\end{landscape}

Figures~\ref{fig_ranking_complexity} and \ref{fig_ranking_knowhow} show the changes in ranking for the drivers estimated from the model. Complexity and collective knowhow go from 1990 to 2016. The ranking of cities only show the top 100 MSAs, and we have highlighted some of the Rust-belt cities.

\newpage
\section{Geographical distribution of collective knowhow in the US}\label{app:map}

\begin{figure}[b!]
	\begin{center}
		\includegraphics[width=0.9\textwidth, trim = 0in 0in 0in 0in]{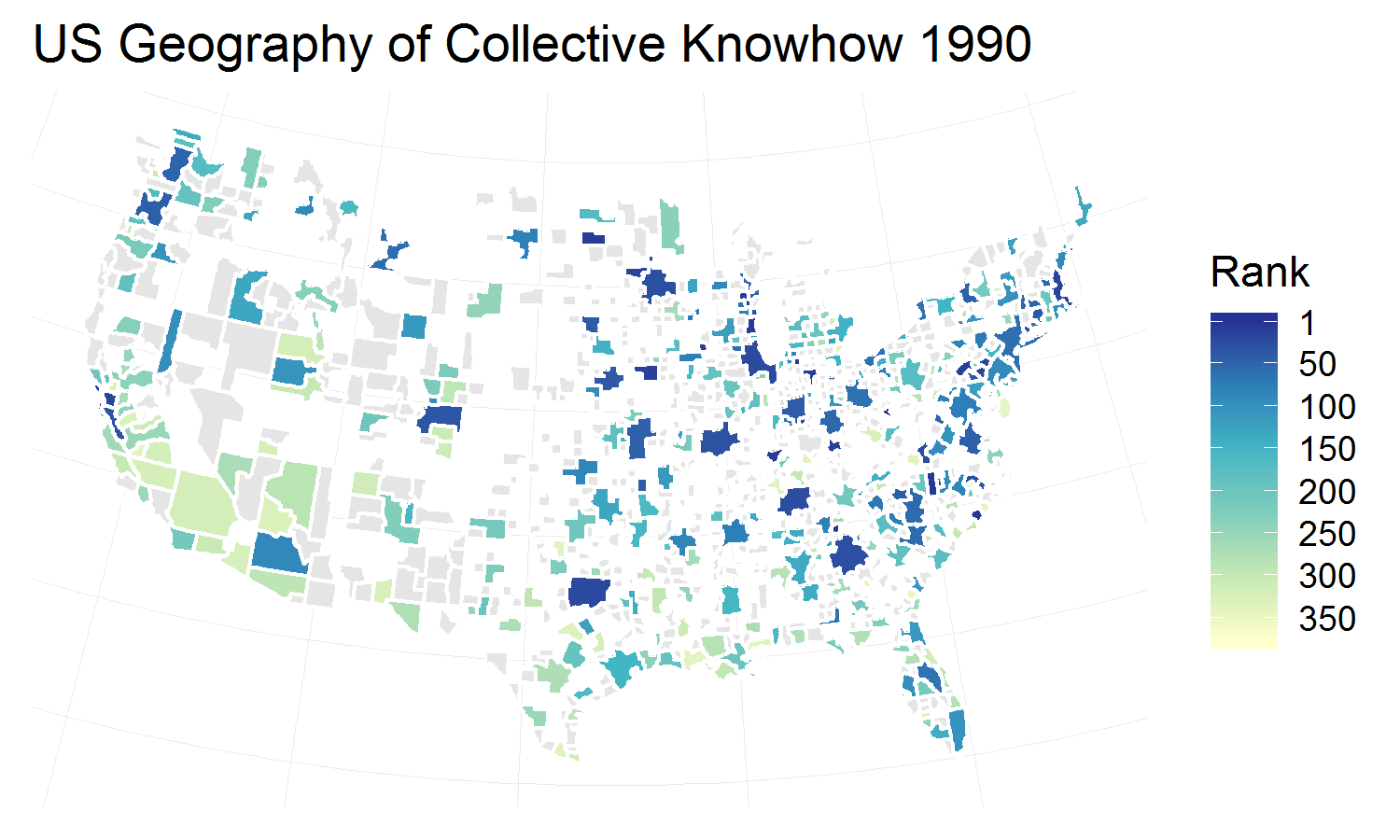}
		\includegraphics[width=0.9\textwidth, trim = 0in 0in 0in 0in]{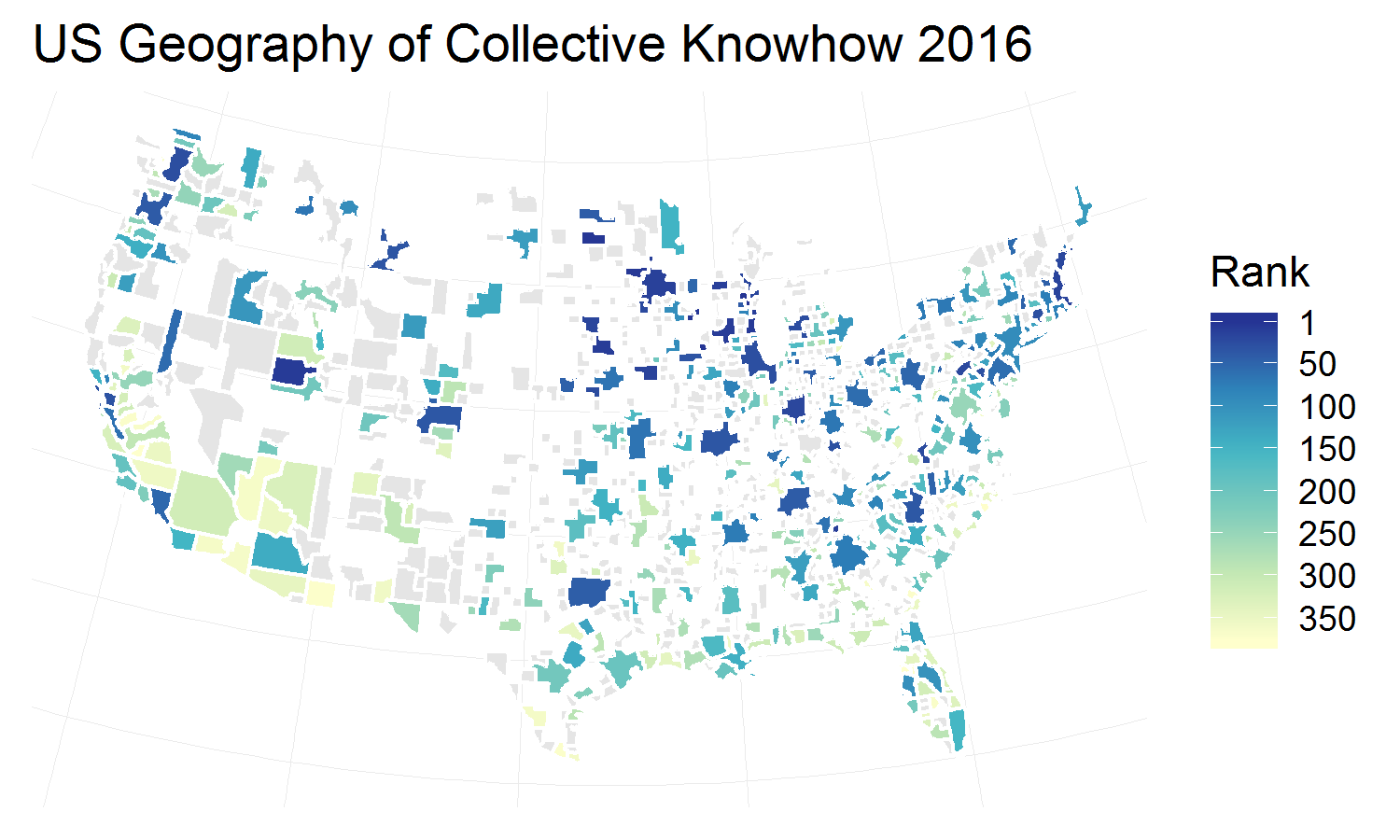}
		\caption{Geographical comparison of the maps in 1990 and 2016 of US Metropolitan Statistical Areas, colored by their ranking in collective knowhow.} 
	\label{Fig_map_comparison}
	\end{center}
\end{figure}

\newpage
\section{Time series of partial elasticities for population size, complexity and collective knowhow}\label{app:Fig_reg_coefficients_long}
\begin{figure}[t!]
	\begin{center}
		\includegraphics[width=0.98\textwidth]{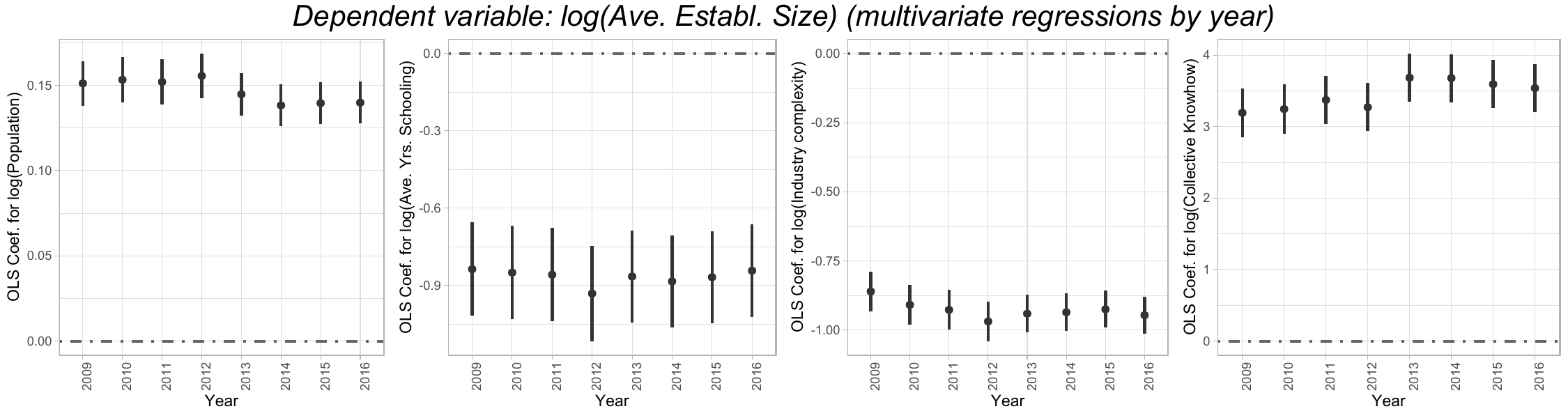}
		\includegraphics[width=0.98\textwidth]{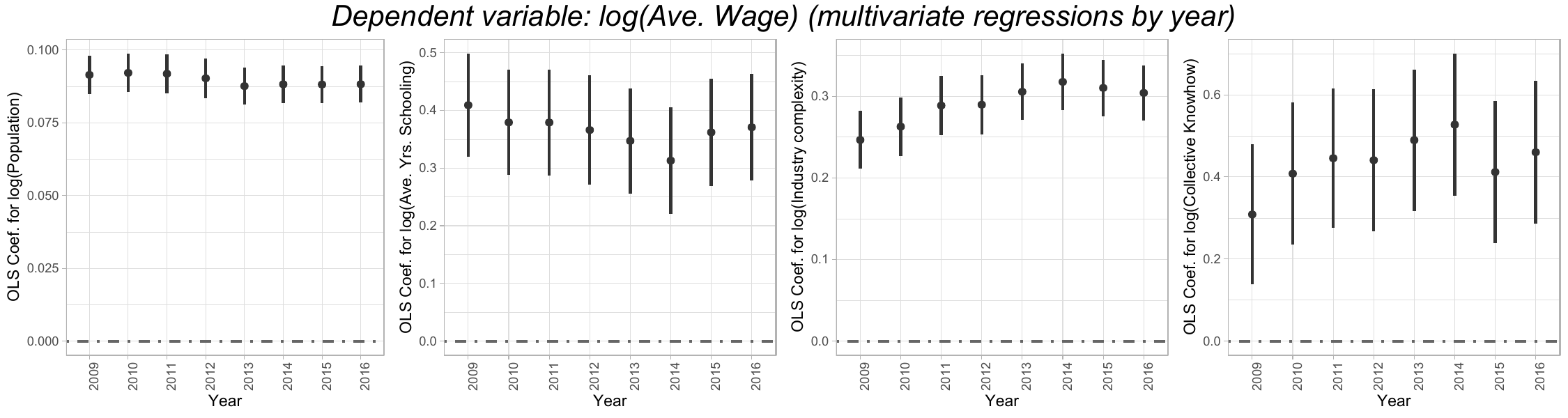}
		\caption{{\bfseries Partial elasticities of firm sizes (above) and wages (below) with respect to regressors.} For each year we carry out a multivariate regression including population, average years of schooling, complexity and collective knowhow (column 6 in Tables~1 and 2 in the main text), and plot the point estimate of the coefficients of such regressions with their corresponding standard error bars.} 
	\label{Fig_reg_coefficients}
	\end{center}
\end{figure}

\begin{figure}[h!]
	\begin{center}
		\includegraphics[width=0.98\textwidth]{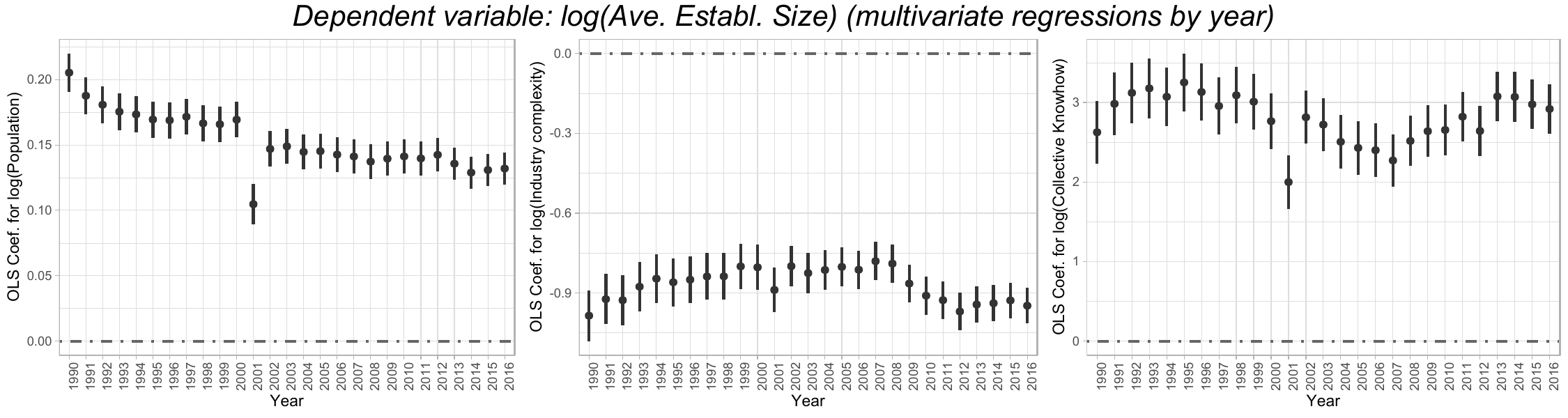}
		\includegraphics[width=0.98\textwidth]{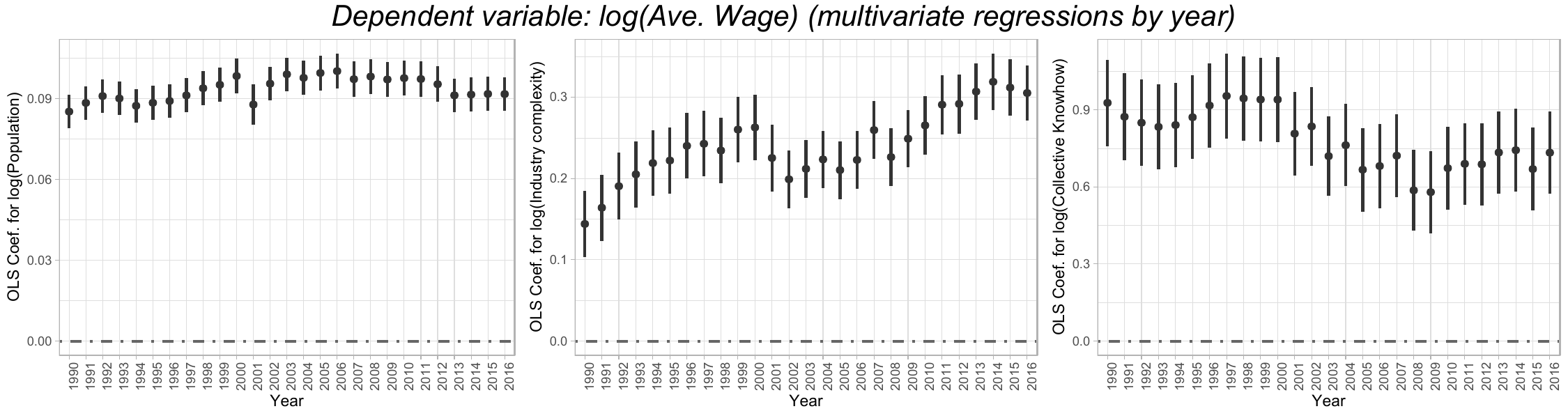}
		\caption{{\bfseries Partial elasticities of firm sizes (above) and wages (below) with respect to regressors.} For each year we carry out a multivariate regression including population, industry complexity and collective knowhow, and plot the point estimate of the coefficients of such regressions with their corresponding standard error bars.} 
	\label{Fig_reg_coefficients_long}
	\end{center}
\end{figure}

Figures~\ref{Fig_reg_coefficients} and \ref{Fig_reg_coefficients_long} show the coefficients of the regressions measuring economic performance (firm size and wages) for different years.

{

\newpage
\section{Regressions of GDP per capita, levels and growth}\label{app:growth_regs}
Even though our theory does not currently incorporate change in time, and therefore we lack a theoretical basis to propose specific growth regressions, we provide here some analysis that correlated the GDP per capita by city with our variables of interest, for both the levels and the growth rate over a 5-year period.

\begin{table}[!t] \centering 
  \caption{Cross-sectional linear regressions for (log) GDP per capita at the city level in 2010 as a function of city (log) population size, (log) of share of workers with a graduate degree, industrial diversity of the city, ECI, and the collective knowhow of city.} 
  \label{tab_gdp_reg} 
\scriptsize
\begin{tabular}{@{\extracolsep{5pt}}lcccccc} 
\\[-1.8ex]\hline 
\hline \\[-1.8ex] 
 & \multicolumn{6}{c}{\textit{Dependent variable:}} \\ 
\cline{2-7} 
\\[-1.8ex] & \multicolumn{6}{c}{`log(GDP per capita)`} \\ 
\\[-1.8ex] & (1) & (2) & (3) & (4) & (5) & (6)\\ 
\hline \\[-1.8ex] 
 `log(Population Size)` & 0.099$^{***}$ &  &  &  &  & 0.053$^{***}$ \\ 
  & t = 8.030 &  &  &  &  & t = 4.648 \\ 
  & & & & & & \\ 
 `log(share w/ Grad. degree)` &  & 0.347$^{***}$ &  &  &  & 0.111$^{***}$ \\ 
  &  & t = 10.623 &  &  &  & t = 3.355 \\ 
  & & & & & & \\ 
 `Industrial Diversity` &  &  & 0.011$^{***}$ &  &  & $-$0.006$^{***}$ \\ 
  &  &  & t = 4.746 &  &  & t = $-$3.115 \\ 
  & & & & & & \\ 
 E.C.I. &  &  &  & 0.097$^{***}$ &  & 0.074$^{***}$ \\ 
  &  &  &  & t = 7.313 &  & t = 6.070 \\ 
  & & & & & & \\ 
 `Collective Knowhow` &  &  &  &  & 3.293$^{***}$ & 3.247$^{***}$ \\ 
  &  &  &  &  & t = 12.301 & t = 12.617 \\ 
  & & & & & & \\ 
 Constant & 9.336$^{***}$ & 11.444$^{***}$ & 10.278$^{***}$ & 10.594$^{***}$ & 10.591$^{***}$ & 10.374$^{***}$ \\ 
  & t = 59.500 & t = 141.157 & t = 152.409 & t = 802.910 & t = 896.306 & t = 64.280 \\ 
  & & & & & & \\ 
\hline \\[-1.8ex] 
Observations & 346 & 345 & 346 & 346 & 346 & 345 \\ 
R$^{2}$ & 0.158 & 0.248 & 0.061 & 0.135 & 0.305 & 0.533 \\ 
Adjusted R$^{2}$ & 0.155 & 0.245 & 0.059 & 0.132 & 0.303 & 0.526 \\ 
\hline 
\hline \\[-1.8ex] 
\textit{Note:}  & \multicolumn{6}{r}{$^{*}$p$<$0.05; $^{**}$p$<$0.01; $^{***}$p$<$0.005} \\ 
\end{tabular} 
\end{table} 
Table~\ref{tab_gdp_reg} shows that our estimate of collective knowhow, when taken alone, has the highest explanatory power (in terms of $R^2$) as compared to all other measures considered (population size, share of skilled workers, industrial diversity, and ECI). When all variables are taken together, they all remain statistically significant, which supports again the idea that collective knowhow captures information that is orthogonal to population size, educational attainment, or industrial composition.

\begin{table}[!t] \centering 
  \caption{Linear regressions for the growth rate of GDP per capita from 2010 to 2015 at the city level as a function of the baseline level of (log) GDP per capita, city (log) population size, (log) of share of workers with a graduate degree, industrial diversity of the city, ECI, and the collective knowhow of city. All independent variables are taken in 2010 and have been standardized to have mean zero and standard deviation equal to one.} 
  \label{tab_gdp_reg_growth} 
\scriptsize
\begin{tabular}{@{\extracolsep{-5pt}}lcccccccc} 
\\[-1.8ex]\hline 
\hline \\[-1.8ex] 
 & \multicolumn{8}{c}{\textit{Dependent variable:}} \\ 
\cline{2-9} 
\\[-1.8ex] & \multicolumn{8}{c}{`GDPpc Growth (5yrs)`} \\ 
\\[-1.8ex] & (1) & (2) & (3) & (4) & (5) & (6) & (7) & (8)\\ 
\hline \\[-1.8ex] 
 `log(GDP per capita)` & 0.082 & 0.023 & 0.108 & 0.051 & 0.132$^{*}$ & $-$0.061 & $-$0.066 & $-$0.126 \\ 
  & t = 1.531 & t = 0.392 & t = 1.743 & t = 0.927 & t = 2.301 & t = $-$0.968 & t = $-$0.870 & t = $-$1.895 \\ 
  & & & & & & & & \\ 
 `log(Population Size)` &  & 0.151$^{**}$ &  &  &  &  & 0.189$^{***}$ & 0.158$^{**}$ \\ 
  &  & t = 2.596 &  &  &  &  & t = 2.852 & t = 2.787 \\ 
  & & & & & & & & \\ 
 `log(share w/ Grad. degree)` &  &  & $-$0.052 &  &  &  & $-$0.063 &  \\ 
  &  &  & t = $-$0.835 &  &  &  & t = $-$0.937 &  \\ 
  & & & & & & & & \\ 
 `Industrial Diversity` &  &  &  & 0.126$^{*}$ &  &  & $-$0.011 &  \\ 
  &  &  &  & t = 2.286 &  &  & t = $-$0.164 &  \\ 
  & & & & & & & & \\ 
 E.C.I. &  &  &  &  & $-$0.136$^{*}$ &  & $-$0.070 &  \\ 
  &  &  &  &  & t = $-$2.364 &  & t = $-$1.016 &  \\ 
  & & & & & & & & \\ 
 `Collective Knowhow` &  &  &  &  &  & 0.260$^{***}$ & 0.242$^{***}$ & 0.265$^{***}$ \\ 
  &  &  &  &  &  & t = 4.118 & t = 3.290 & t = 4.241 \\ 
  & & & & & & & & \\ 
 Constant & $-$0.000 & $-$0.000 & $-$0.000 & $-$0.000 & $-$0.000 & $-$0.000 & 0.000 & 0.000 \\ 
  & t = $-$0.000 & t = $-$0.000 & t = $-$0.000 & t = $-$0.000 & t = $-$0.000 & t = $-$0.000 & t = 0.000 & t = 0.000 \\ 
  & & & & & & & & \\ 
\hline \\[-1.8ex] 
Observations & 345 & 345 & 345 & 345 & 345 & 345 & 345 & 345 \\ 
R$^{2}$ & 0.007 & 0.026 & 0.009 & 0.022 & 0.023 & 0.054 & 0.084 & 0.075 \\ 
Adjusted R$^{2}$ & 0.004 & 0.020 & 0.003 & 0.016 & 0.017 & 0.048 & 0.068 & 0.067 \\ 
\hline 
\hline \\[-1.8ex] 
\textit{Note:}  & \multicolumn{8}{r}{$^{*}$p$<$0.05; $^{**}$p$<$0.01; $^{***}$p$<$0.005} \\ 
\end{tabular} 
\end{table}
Equally interesting are the results from the growth regressions in Table~\ref{tab_gdp_reg_growth}. In the regressions, the dependent variable is $\ln(y_{t+5}/y_t)$, where $y$ is GDP per capita and $t=2010$. In addition to the independent variables considered in Table~\ref{tab_gdp_reg}, we also include the (log) level of GDP per capita in 2010, to capture the effect of reversions to the mean. All independent variables are taken also in 2010. Furthermore, to add clarity to the interpretation and comparison of the regression coefficients, all variables have been standardized. 

As can be observed from the results, predicting the growth rate in GDP per capita is a hard problem and our model specifications are only able to explain a small portion of the variation in the data. Still, some interesting observations are worth noting. As a start, the coefficient for GDP per capita in the regressions shown in Table~\ref{tab_gdp_reg_growth} are all positive except when our estimate for the collective knowhow of the city is added to the regression (columns 6, 7 and 8). That is, predicting GDP growth shows evidence of reversion to the mean only when measure for collective knowhow is considered. This observation suggests something crucial about our measure of collective knowhow. We list the following as crucial facts. First, we know the level of GDP per capita at the baseline year already captures a lot of the information for why a city is wealthy, and why it may become wealthier. Second, we know from Table~\ref{tab_gdp_reg} that GDP per capita is highly correlated with collective knowhow. Third, we observe that collective knowhow does not drop from regressions 6, 7 or 8 in Table~\ref{tab_gdp_reg_growth} (i.e., its statistical significance does not go away) even when all variables are taken (implying that collective knowhow carries orthogonal information to GDP per capita, despite its high correlation with the latter). Fourth, we note that the magnitude of its effect is the largest among all the variables (recall, all the variables have been standardized) and is stable across specifications. And fifth, the contribution to explaining the variation in GDP growth in terms of $R^2$ accounted is also the largest. Taken together, all these facts are highly suggestive evidence that supports our interpretation of our estimate of collective knowhow as a true driver of economic growth.

}

%% file: NHB_10001_industries_excluded.tex
\begin{table}[!htbp] \centering 
  \caption{Industries excluded from analysis due to the fact that their appearance in cities is determined by government or mainly driven by availability of demand as opposed to availability of capacities for supply.} 
  \label{tab_inds_excluded} 
\scriptsize 
\begin{tabular}{@{\extracolsep{0pt}} cl} 
\\[-1.8ex]\hline 
\hline \\[-1.8ex] 
NAICS & Title \\ 
\hline \\[-1.8ex] 
113 & Forestry and logging \\ 
114 & Fishing, hunting and trapping \\ 
115 & Agriculture and forestry support activities \\ 
211 & Oil and gas extraction \\ 
212 & Mining, except oil and gas \\ 
213 & Support activities for mining \\ 
447 & Gasoline stations \\ 
482 & Rail transportation \\ 
483 & Water transportation \\ 
486 & Pipeline transportation \\ 
487 & Scenic and sightseeing transportation \\ 
491 & Postal service \\ 
516 & Internet publishing and broadcasting \\ 
521 & Monetary authorities - central bank \\ 
921 & Executive, legislative and general government \\ 
922 & Justice, public order, and safety activities \\ 
923 & Administration of human resource programs \\ 
924 & Administration of environmental programs \\ 
925 & Community and housing program administration \\ 
926 & Administration of economic programs \\ 
927 & Space research and technology \\ 
928 & National security and international affairs \\ 
999 & Unclassified \\ 
\hline \\[-1.8ex] 
\end{tabular} 
\end{table} 

%% file: NHB_10001_industries_included.tex
\begin{table}[!htbp] \centering 
  \caption{Industries included in our analysis due to the fact that their appearance in cities is mainly supply-driven.} 
  \label{tab_inds_included} 
\scriptsize 
\begin{tabular}{@{\extracolsep{0pt}} cl|cl} 
\\[-1.8ex]\hline 
\hline \\[-1.8ex] 
NAICS & Title & NAICS & Title \\ 
\hline \\[-1.8ex] 
111 & Crop production & 453 & Miscellaneous store retailers \\ 
112 & Animal production and aquaculture & 454 & Nonstore retailers \\ 
221 & Utilities & 481 & Air transportation \\ 
236 & Construction of buildings & 484 & Truck transportation \\ 
237 & Heavy and civil engineering construction & 485 & Transit and ground passenger transportation \\ 
238 & Specialty trade contractors & 488 & Support activities for transportation \\ 
311 & Food manufacturing & 492 & Couriers and messengers \\ 
312 & Beverage and tobacco product manufacturing & 493 & Warehousing and storage \\ 
313 & Textile mills & 511 & Publishing industries, except internet \\ 
314 & Textile product mills & 512 & Motion picture and sound recording industries \\ 
315 & Apparel manufacturing & 515 & Broadcasting, except internet \\ 
316 & Leather and allied product manufacturing & 517 & Telecommunications \\ 
321 & Wood product manufacturing & 518 & Data processing, hosting and related services \\ 
322 & Paper manufacturing & 519 & Other information services \\ 
323 & Printing and related support activities & 522 & Credit intermediation and related activities \\ 
324 & Petroleum and coal products manufacturing & 523 & Securities, commodity contracts, investments \\ 
325 & Chemical manufacturing & 524 & Insurance carriers and related activities \\ 
326 & Plastics and rubber products manufacturing & 525 & Funds, trusts, and other financial vehicles \\ 
327 & Nonmetallic mineral product manufacturing & 531 & Real estate \\ 
331 & Primary metal manufacturing & 532 & Rental and leasing services \\ 
332 & Fabricated metal product manufacturing & 533 & Lessors of nonfinancial intangible assets \\ 
333 & Machinery manufacturing & 541 & Professional and technical services \\ 
334 & Computer and electronic product manufacturing & 551 & Management of companies and enterprises \\ 
335 & Electrical equipment and appliance mfg. & 561 & Administrative and support services \\ 
336 & Transportation equipment manufacturing & 562 & Waste management and remediation services \\ 
337 & Furniture and related product manufacturing & 611 & Educational services \\ 
339 & Miscellaneous manufacturing & 621 & Ambulatory health care services \\ 
423 & Merchant wholesalers, durable goods & 622 & Hospitals \\ 
424 & Merchant wholesalers, nondurable goods & 623 & Nursing and residential care facilities \\ 
425 & Electronic markets and agents and brokers & 624 & Social assistance \\ 
441 & Motor vehicle and parts dealers & 711 & Performing arts and spectator sports \\ 
442 & Furniture and home furnishings stores & 712 & Museums, historical sites, zoos, and parks \\ 
443 & Electronics and appliance stores & 713 & Amusements, gambling, and recreation \\ 
444 & Building material and garden supply stores & 721 & Accommodation \\ 
445 & Food and beverage stores & 722 & Food services and drinking places \\ 
446 & Health and personal care stores & 811 & Repair and maintenance \\ 
448 & Clothing and clothing accessories stores & 812 & Personal and laundry services \\ 
451 & Sports, hobby, music instrument, book stores & 813 & Membership associations and organizations \\ 
452 & General merchandise stores & 814 & Private households \\ 
\hline \\[-1.8ex] 
\end{tabular} 
\end{table}